\documentclass[twocolumn,final]{revtex4}
\usepackage{amsmath}
\usepackage{mathtools}
\usepackage{amssymb}
\usepackage{color}
\usepackage{graphicx}
\usepackage{hyperref}
\usepackage{url}
\newcommand{\al}{\alpha}
\newcommand{\be}{\beta}
\newcommand{\fr}{\frac}
\newcommand{\Ga}{\Gamma}
\newcommand{\ga}{\gamma}
\newcommand{\lf}{\left}
\newcommand{\rg}{\right}
\newcommand{\si}{\sigma}

\newcommand{\LN}{\text{LN}}
\newcommand{\IGa}{\text{IGa}}
\newcommand{\GIGa}{\text{GIGa}}
\newcommand{\GGa}{\text{GGa}}

\begin{document}
\title{Distribution of Human Response Times}
\author{Tao Ma}
\affiliation{Department of Physics, University of Cincinnati, Cincinnati, OH 45221-0011}
\author{John G. Holden}
\email{john.holden@uc.edu}
\affiliation{CAP Center for Cognition, Action, and Perception, Department of Psychology, University of Cincinnati, Cincinnati, OH 45221-0376}
\author{R.A. Serota}
\email{serota@ucmail.uc.edu}
\affiliation{Department of Physics, University of Cincinnati, Cincinnati, OH 45221-0011}
\date{\today}
\begin{abstract}
We demonstrate that distributions of human response times have power-law tails and, among closed-form distributions, are best fit by the generalized inverse gamma distribution. We speculate that the task difficulty tracks the half-width of the distribution and show that it is related to the exponent of the power-law tail. 
\end{abstract}
\maketitle

\section{Introduction}

Human response time (RT) is defined as the time delay between a signal and the starting point of human action. For example, one can measure time interval from a word appearing on a computer screen to when a participant pushes a keyboard button to indicate his or her response. Two well established empirical facts of RT are the power law tails of RT distributions \cite{holden2012} and 1/f noise of RT time series \cite{Van2003,Van2005,Kello2010,Wagenmakers2004}, to which any theoretical description must conform. 

The generalized inverse gamma (GIGa) function (Appendix \ref{GIGa_Scale}) belongs to a family of distributions (Appendix \ref{GIGa_LN}), which includes inverse gamma (IGa), lognormal (LN), gamma (Ga) and generalized gamma (GGa). The remarkable property of GIGa is its power-law tail; for a general three-parameter case, the power-law exponent is given by the negative $1+\al\ga$, so that $\GIGa(x; \al,\be,\ga)\propto x^{-1-\al\ga}$, $x\rightarrow \infty$. GIGa emerges as a steady state distribution in a number of systems, from a network model of economy, \cite{ma2013distribution} to ontogenetic mass growth, \cite{west2012} to stock volatility \cite{ma13ST}. This common feature can be traced to a birth-death phenomenological model subject to stochastic perturbations (Appendix \ref{Birth_Death}). 

Here we argue that among closed form distributions the GIGa best describes RT distribution. GIGa has a natural scale parameter, which determines the onset of the power law tail, and two shape parameters, which determine the exponent of the tail. As such, our argument is an extension of previous approaches, such as ``cocktail" model, \cite{holden2012} which effectively contains shape and scale parameters as well. Furthermore, we speculate that the difficulty of a cognitive task tracks the half-width of the RT distribution and discuss it within the GIGa framework. 

Our numerical analysis is performed on the following data (explained in text): ELP (English Lexicon Project), HE (Hick's Experiments) and LDT (Lexical Decision Time). Two key features distinguish our approach. First, in addition to usual individual participant fitting, we perform distribution fitting on combined participants' data. While in line with individual fitting, this creates considerably less noisy sets of data. Second, we develop a procedure for fitting the tails of the distribution directly (Appendix \ref{Tail_Power}), which unequivocally proves the existence of power law tails. 

This paper is organized as follows. In Section II, we provide description of the experimental setup and data acquisition. In Section III, we conduct log-log tail fitting and RT distribution fitting with GIGa. In Section IV, we conclude with the discussion of task difficulty.

\section{Data acquisition}
\subsection{Data sources and descritpion}
ELP data is from the English Lexicon Project \cite{Balota2007,ELP_website}. HE and LDT data was collected under the supervision of J. G. Holden. 

\subsubsection{ELP}
ELP (English Lexicon Project) studies pronunciation latencies to visually presented words; participants sampled from six different Universities. \cite{Balota2007,ELP_website}. 

Data: Two sessions, 470 participants each: session 1 (ELP1), 1500 trials; session 2 (ELP2), 1030 trials. 

\subsubsection{HE}
HE (Hick's Choice RT Experiment) -- given a stimulus selected from a finite set of stimuli, participants try to respond with an action from a set of actions corresponding to this set of stimuli. Original HE is described in \cite{hick1952}.

Data: 11 participants completed 1440 trials of 2, 4, 6, 8, and 10 options, approximately 16 000 combined datapoints for each condition. 

\subsubsection{LDT}
LDT (Lexical Decision Time). 

Data: Three groups 60 participants completed 100 word and 100 nonword trials of 1, 2, and 4 word LDT respectively,
only the correct word trials are depicted, approximately 6000 datapoints for each group. 

\subsection{Data preprocessing}
To enhance our efforts to understand the distribution's tail behavior, we combined all participants' data from each experiment into a single distribution. 

\section{Data analysis}
\subsection{Tail fitting}

\begin{figure}[htp]
\centering
\includegraphics[width=0.23\textwidth]{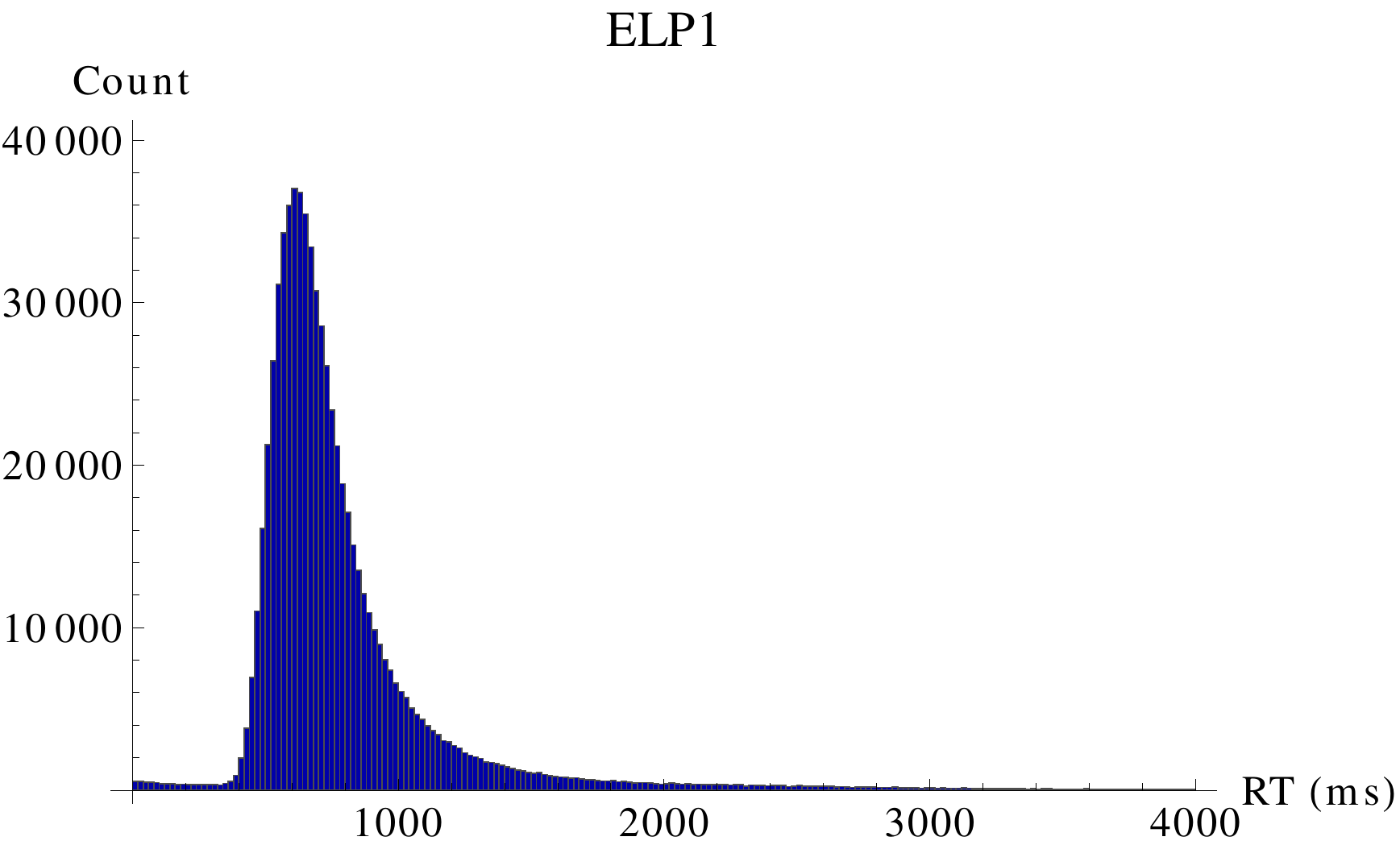}
\includegraphics[width=0.23\textwidth]{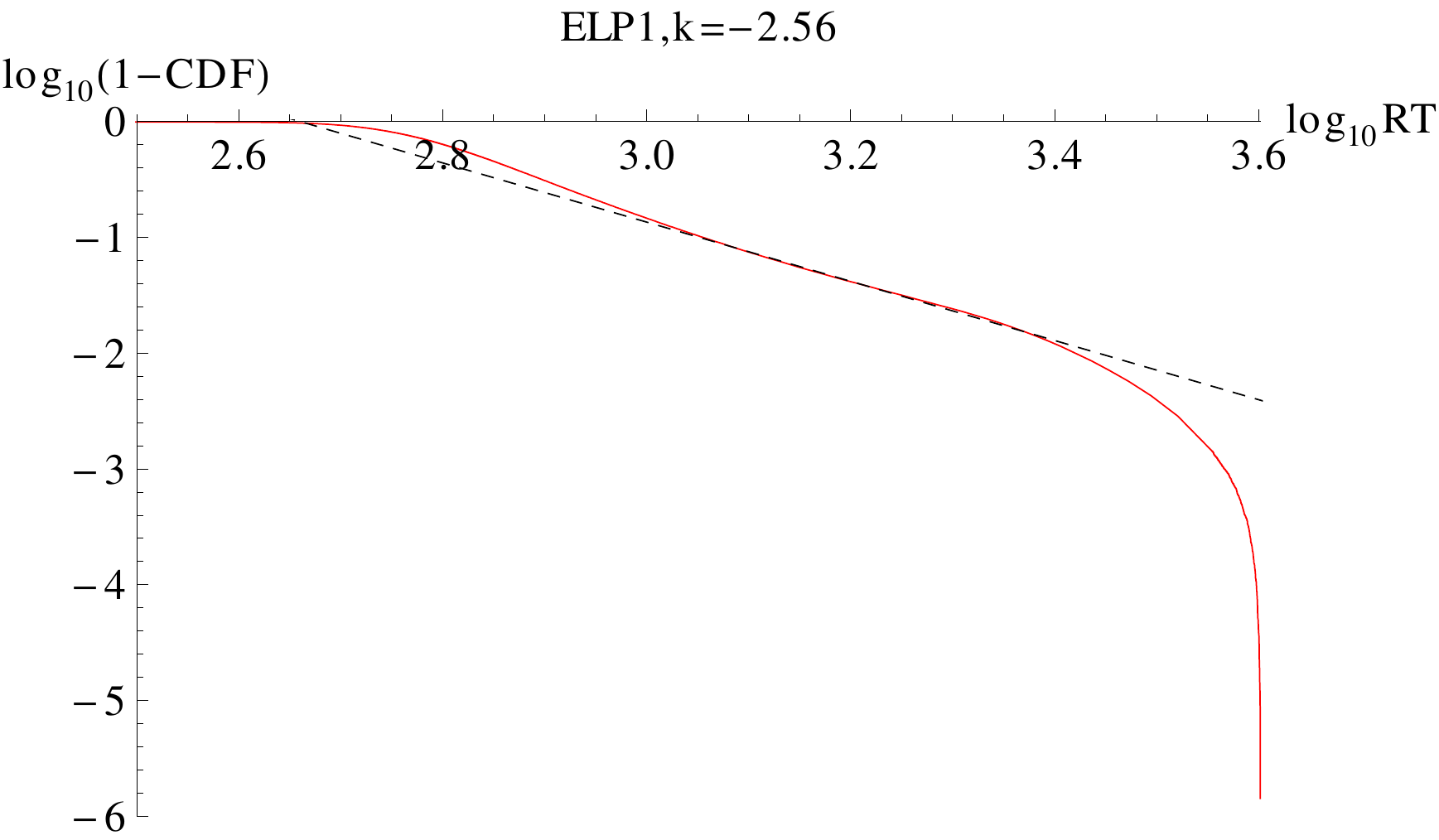}\\
\includegraphics[width=0.23\textwidth]{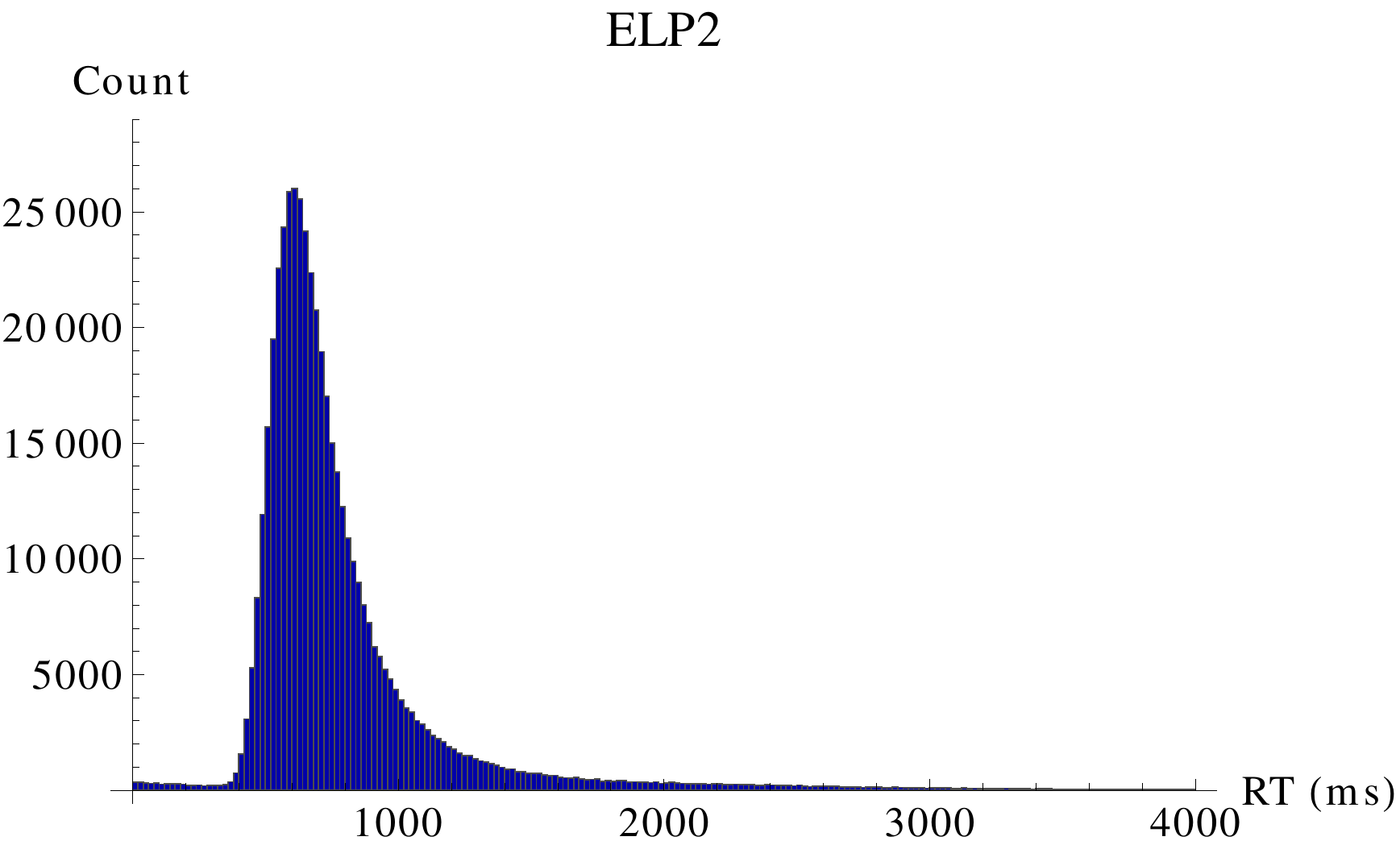}
\includegraphics[width=0.23\textwidth]{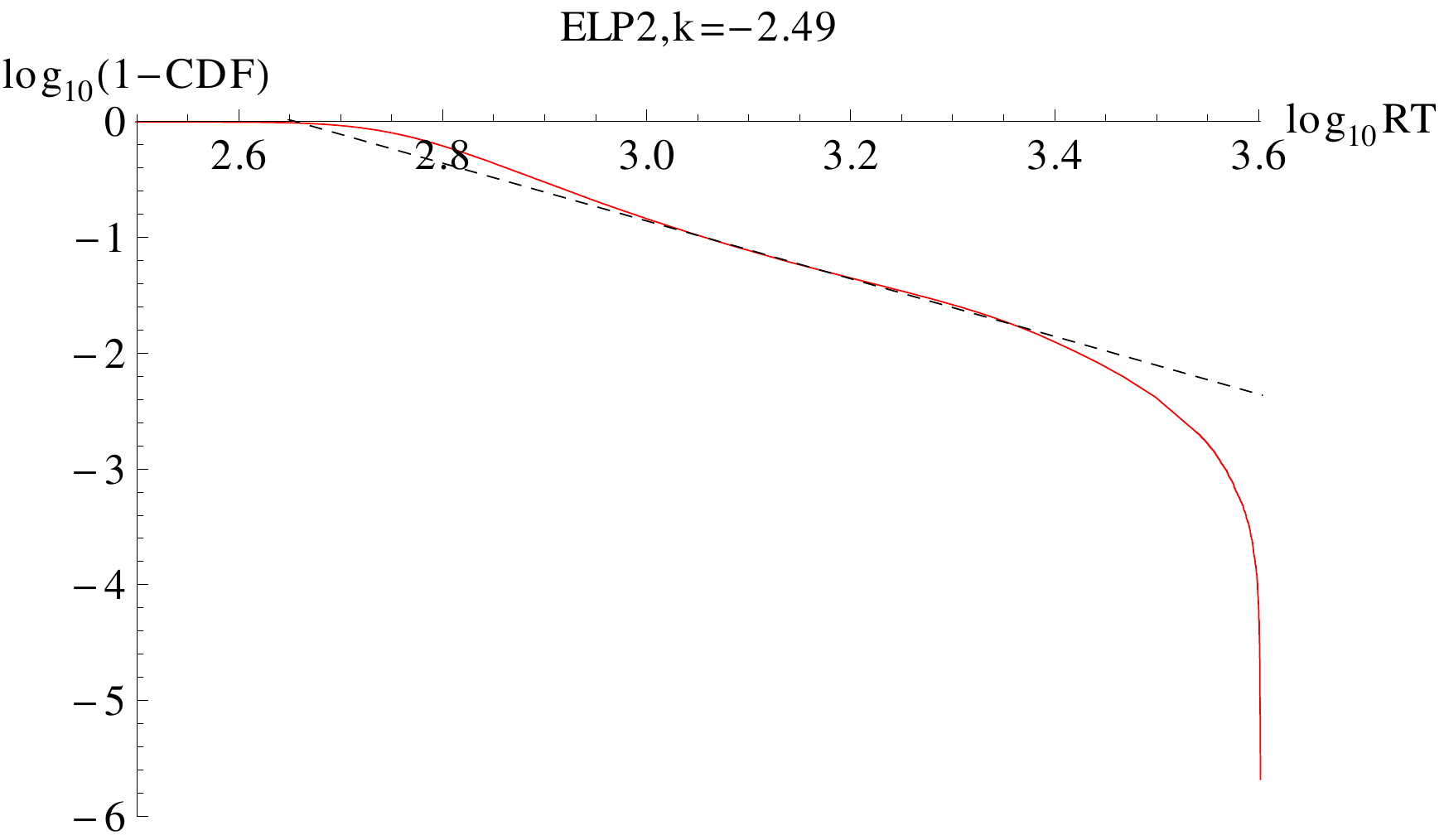}
\caption{Histogram and log-log plot of ELP.}
\label{RT:fig:loglog_ELP}
\end{figure}

\begin{figure}[htp]
\centering
\includegraphics[width=0.23\textwidth]{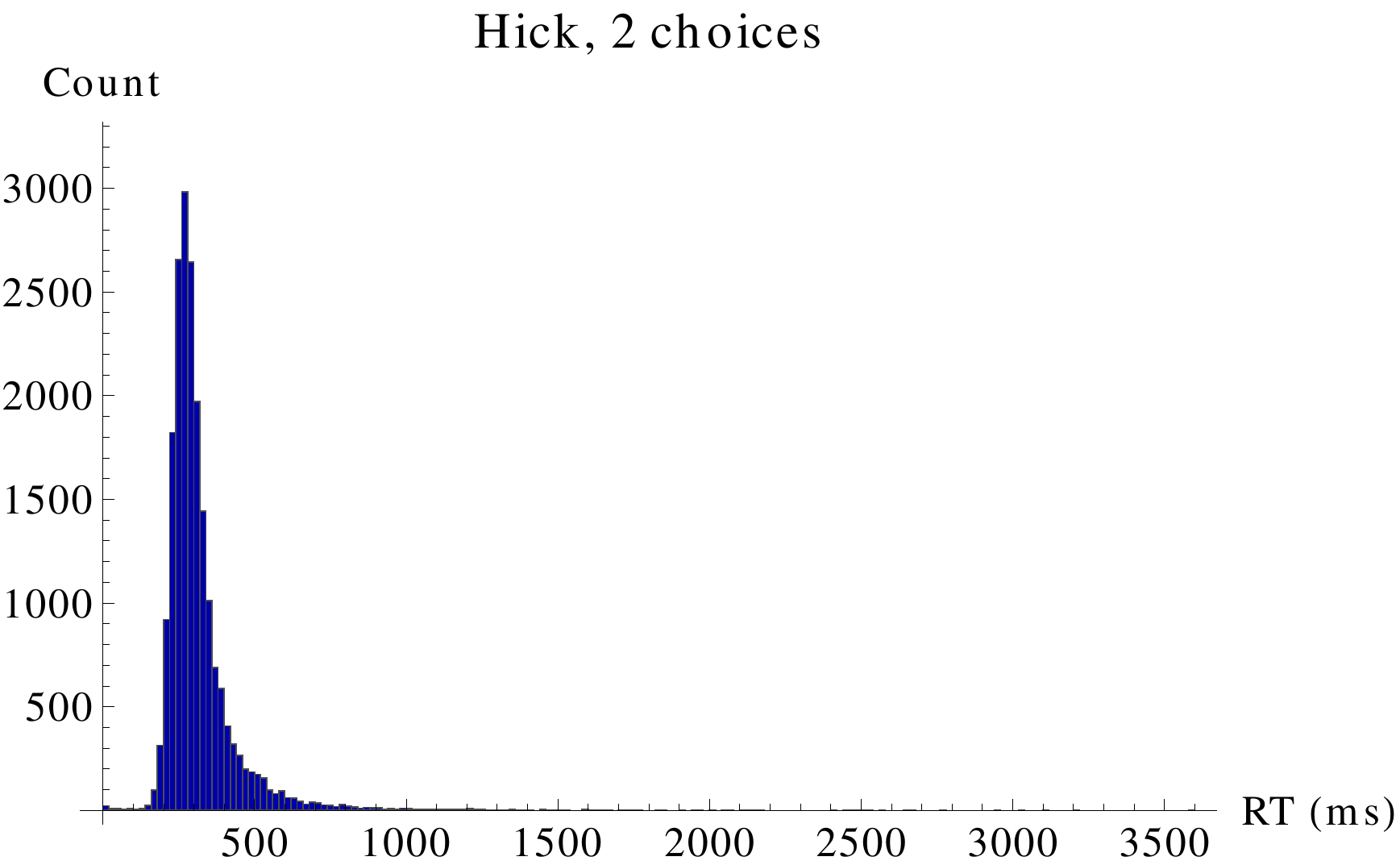}
\includegraphics[width=0.23\textwidth]{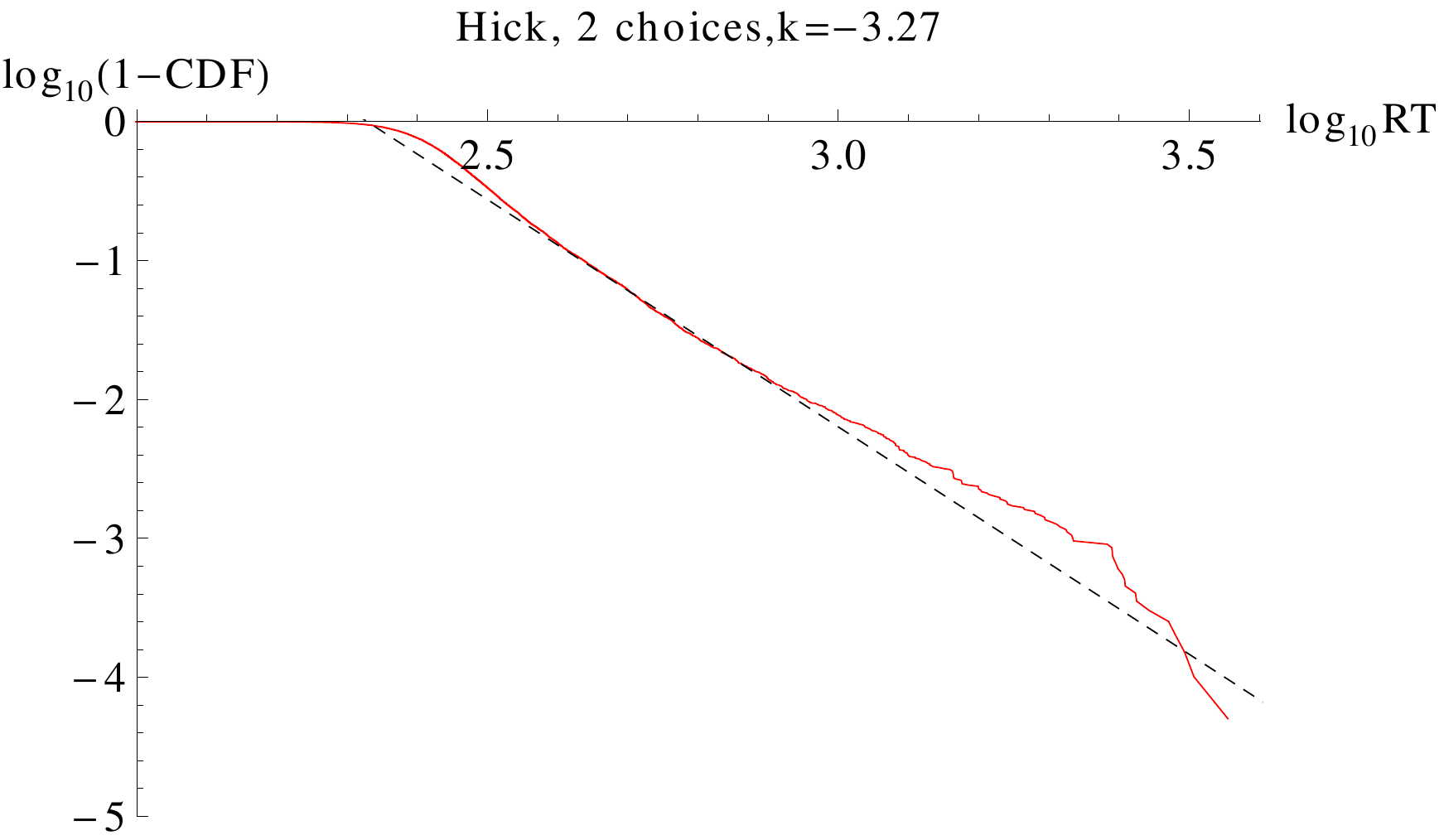}\\
\includegraphics[width=0.23\textwidth]{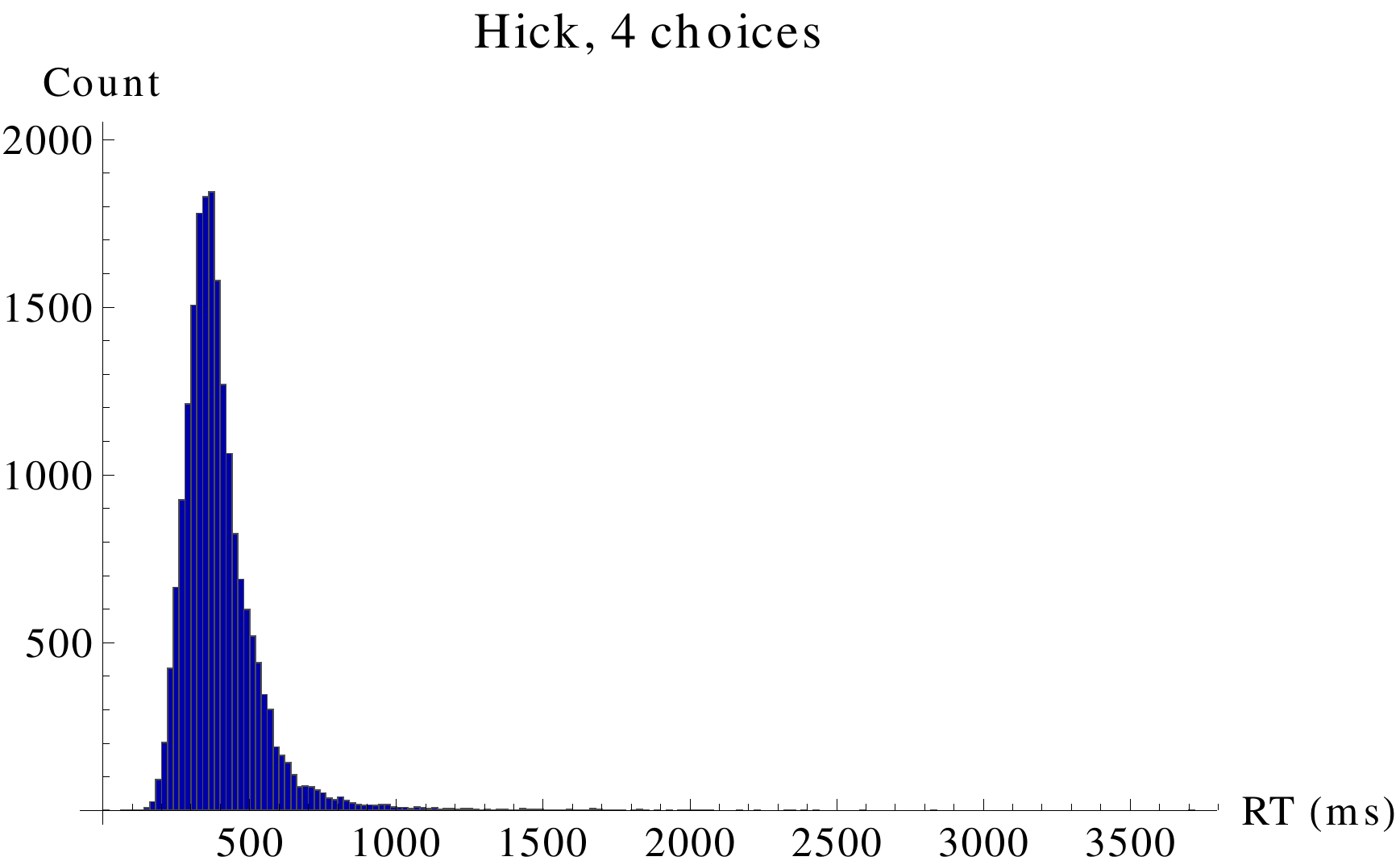}
\includegraphics[width=0.23\textwidth]{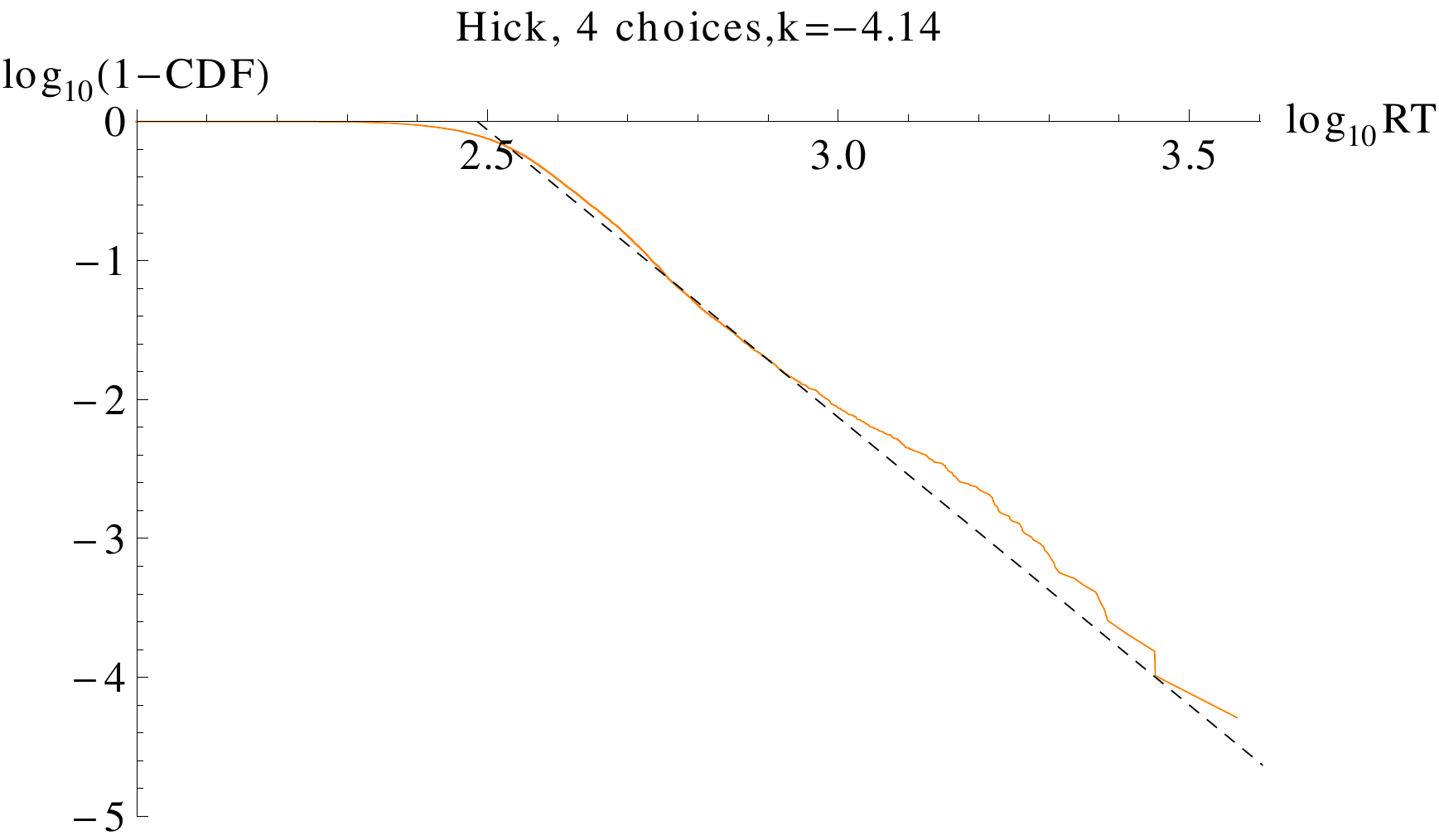}\\
\includegraphics[width=0.23\textwidth]{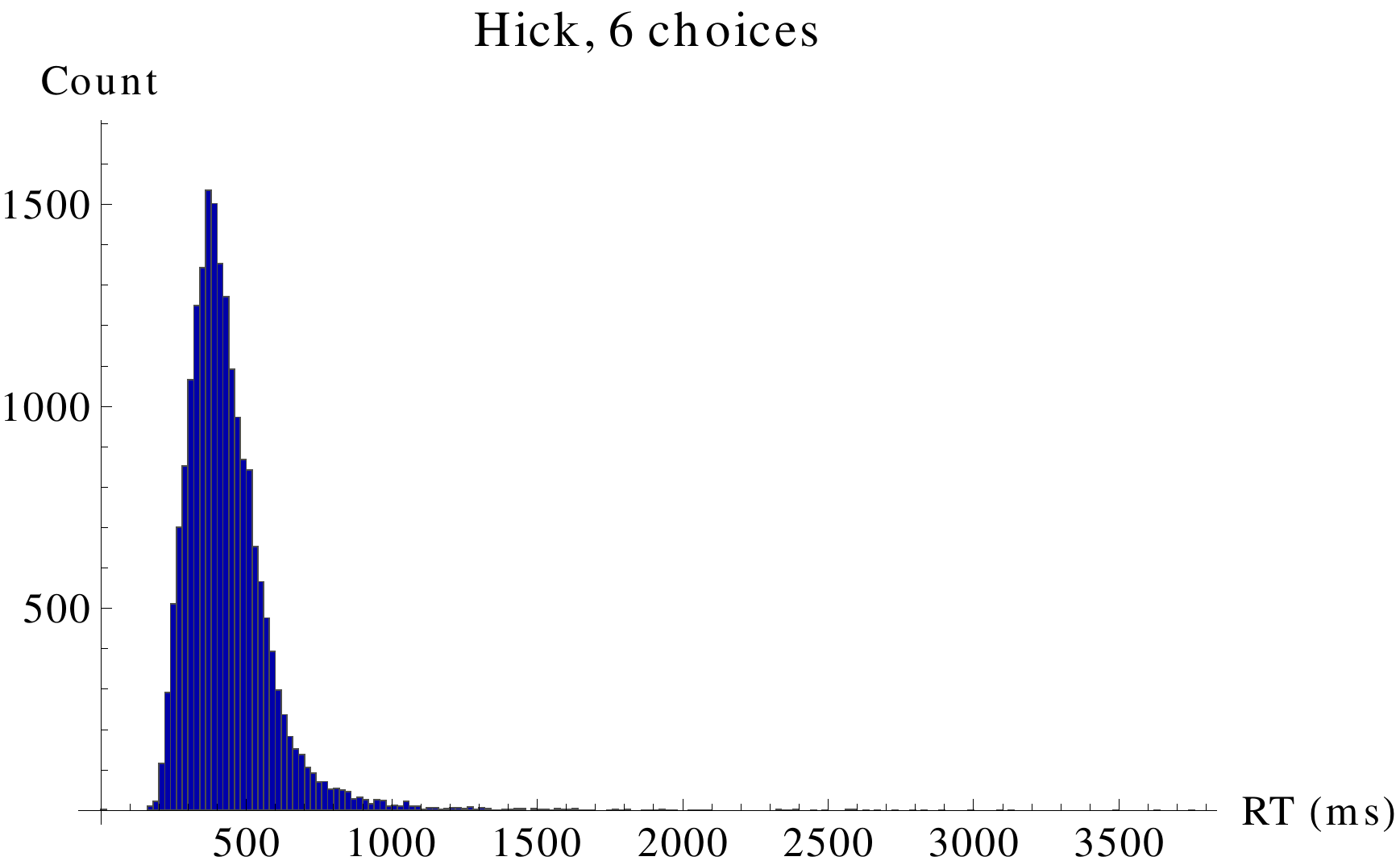}
\includegraphics[width=0.23\textwidth]{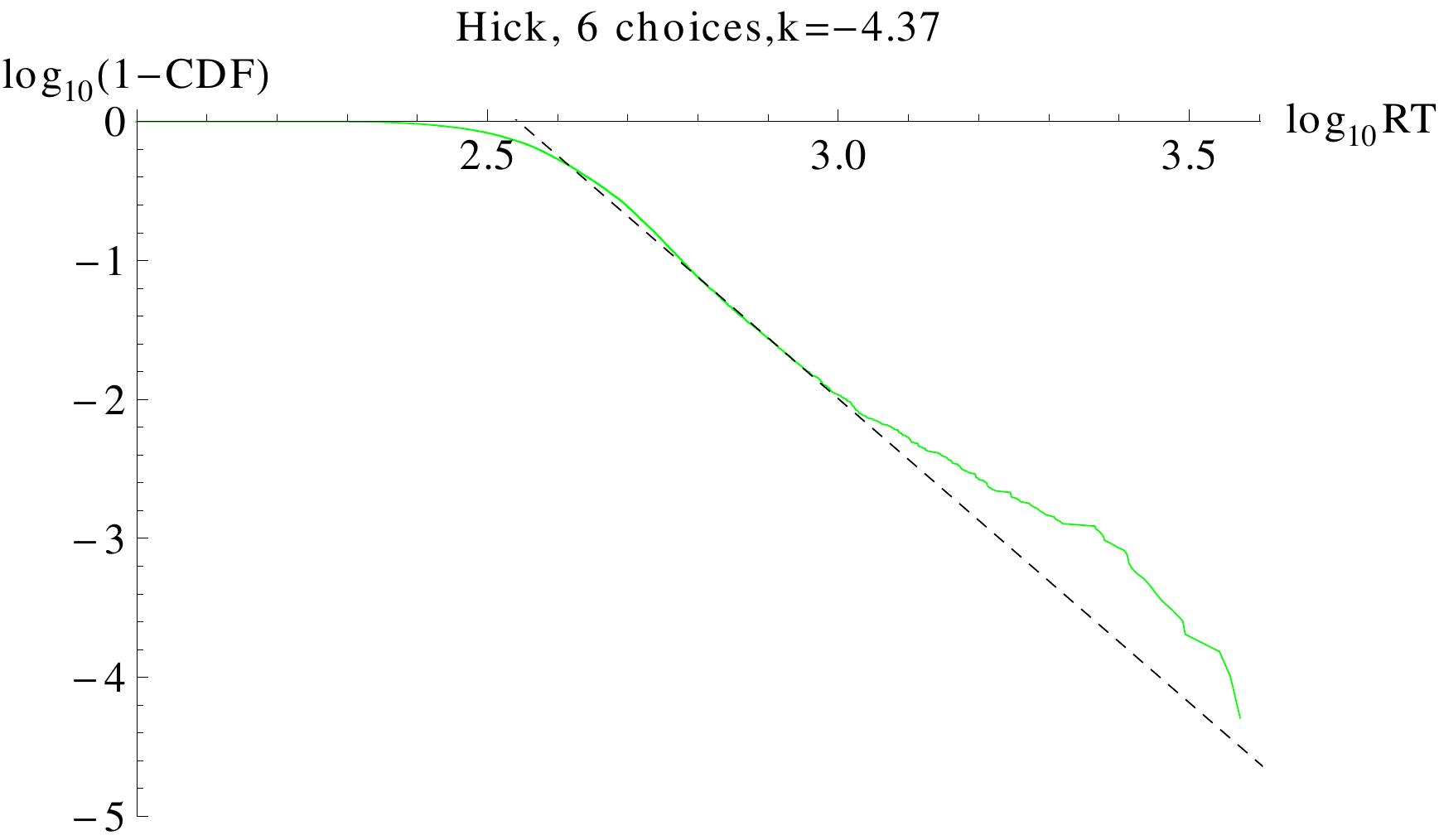}\\
\includegraphics[width=0.23\textwidth]{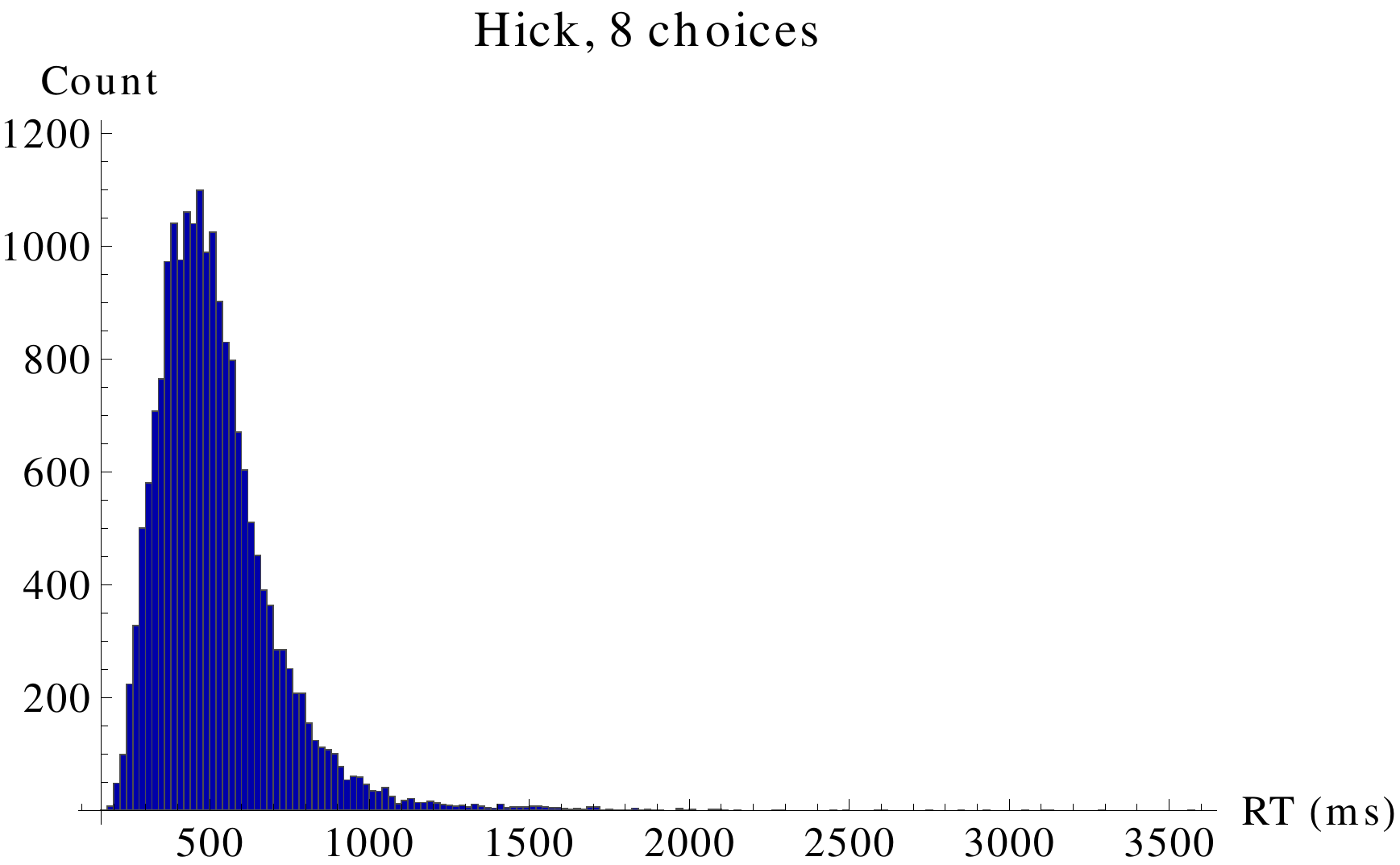}
\includegraphics[width=0.23\textwidth]{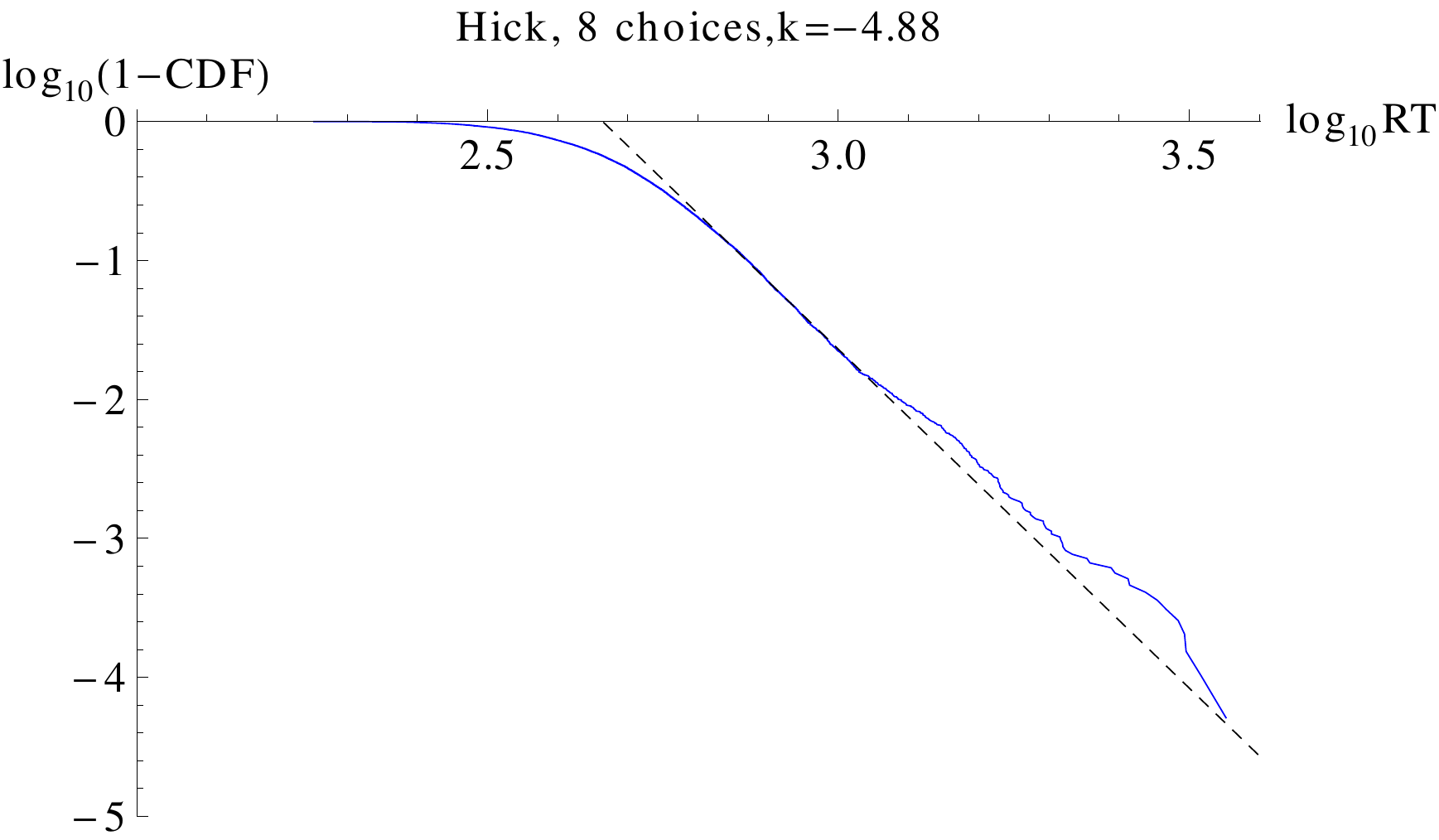}
\caption{Histogram and log-log plot of Hick's experiment.}
\label{RT:fig:loglog_Hick}
\end{figure}

\begin{figure}[htp]
\centering
\includegraphics[width=0.23\textwidth]{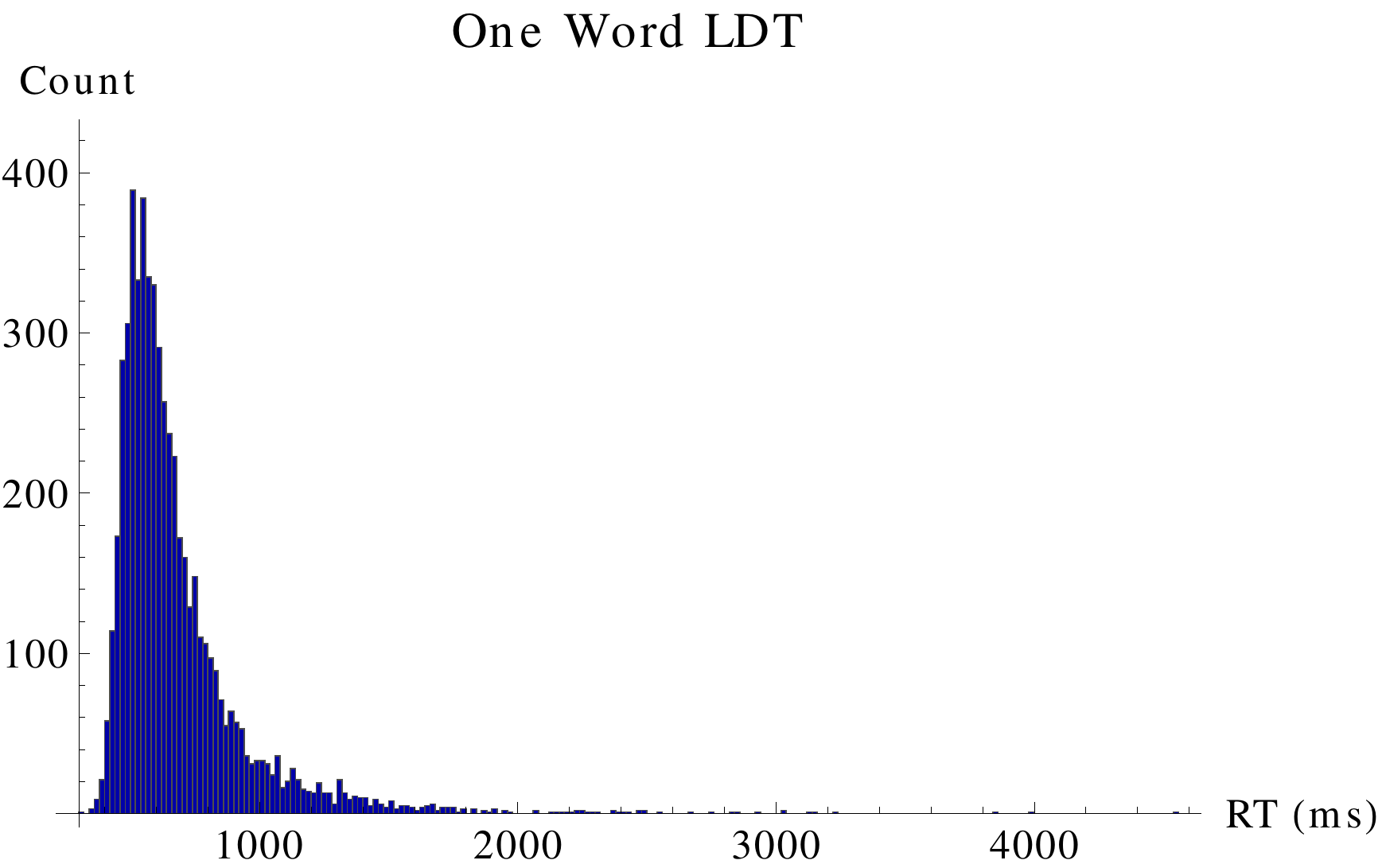}
\includegraphics[width=0.23\textwidth]{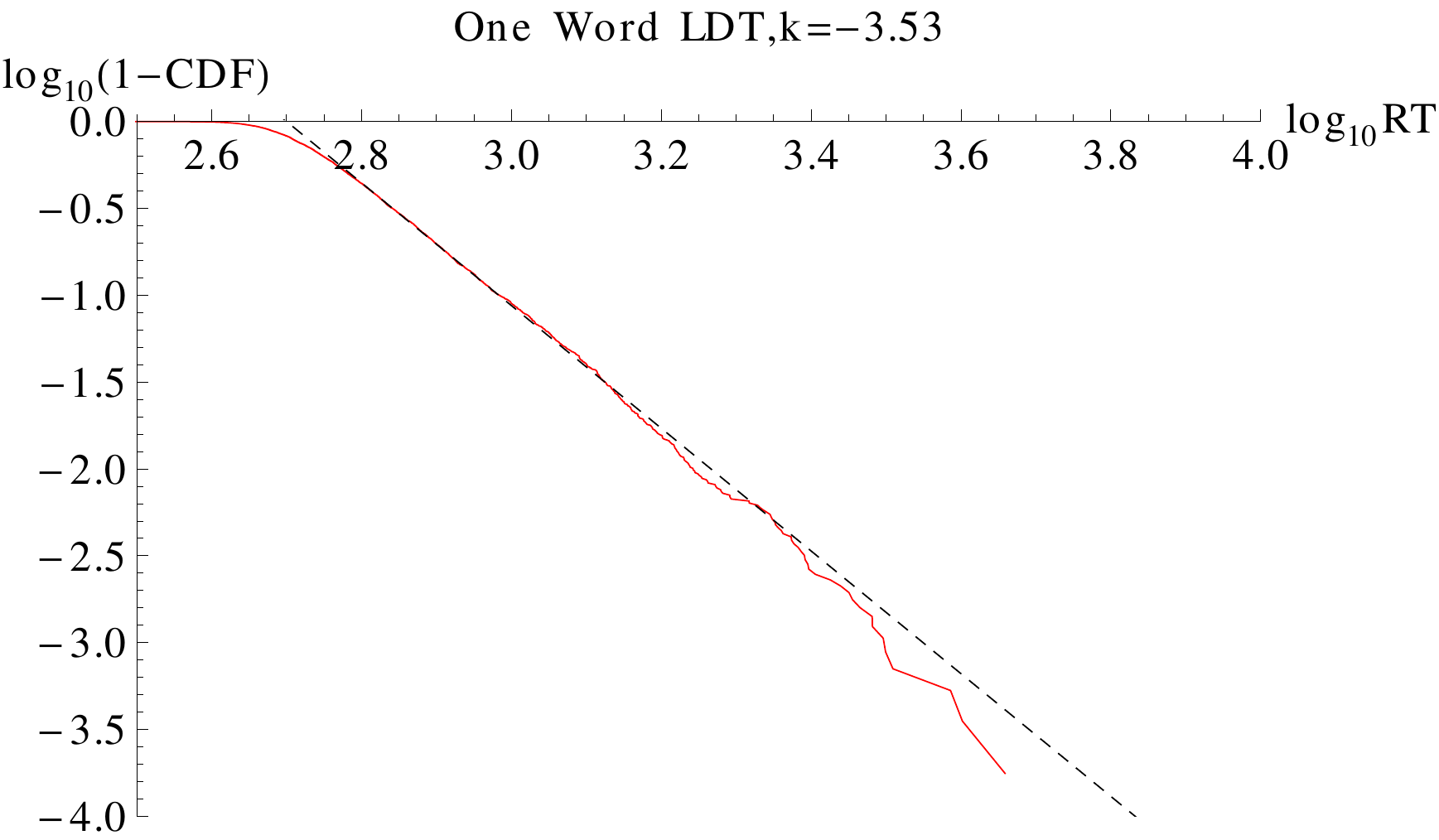}\\
\includegraphics[width=0.23\textwidth]{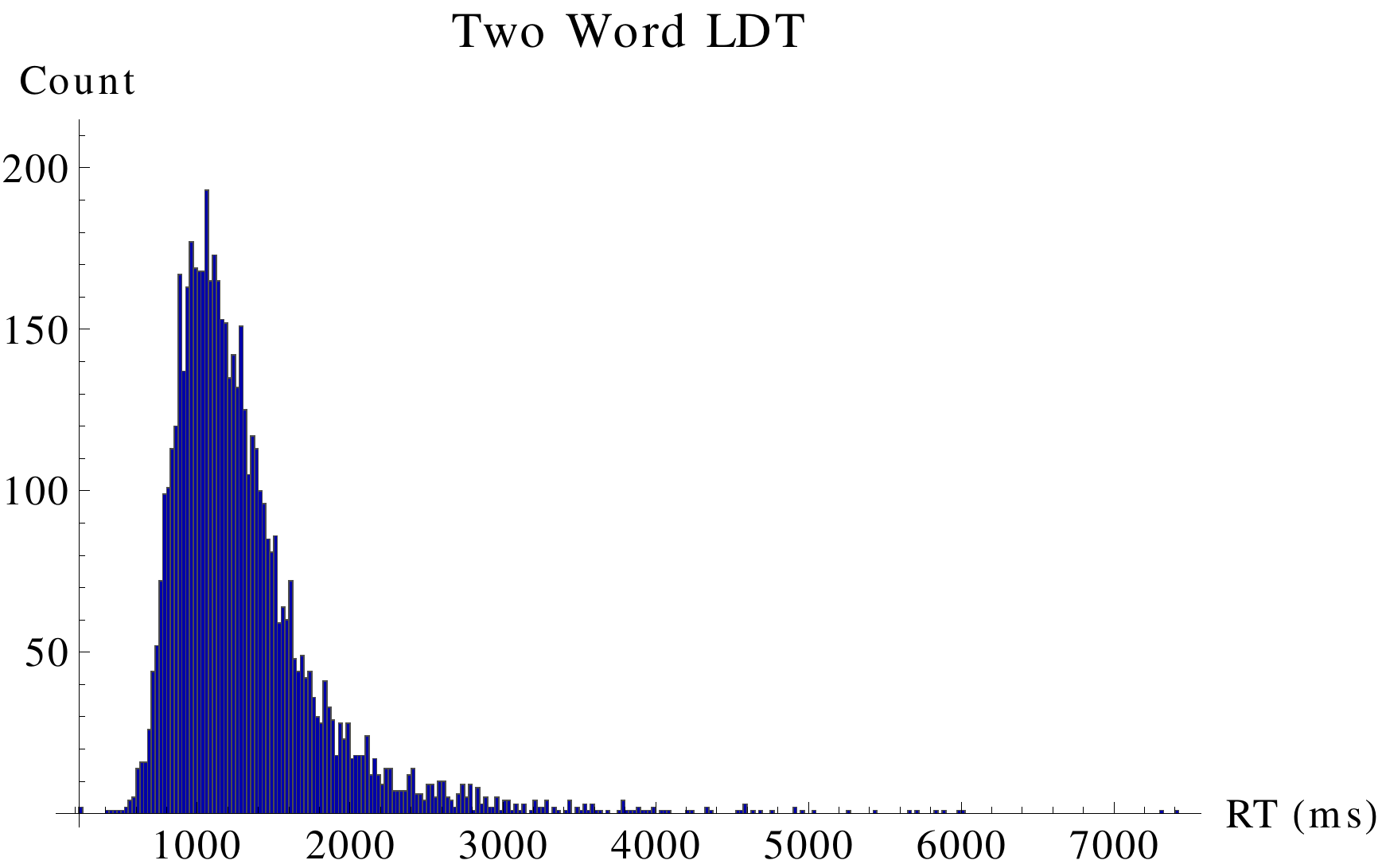}
\includegraphics[width=0.23\textwidth]{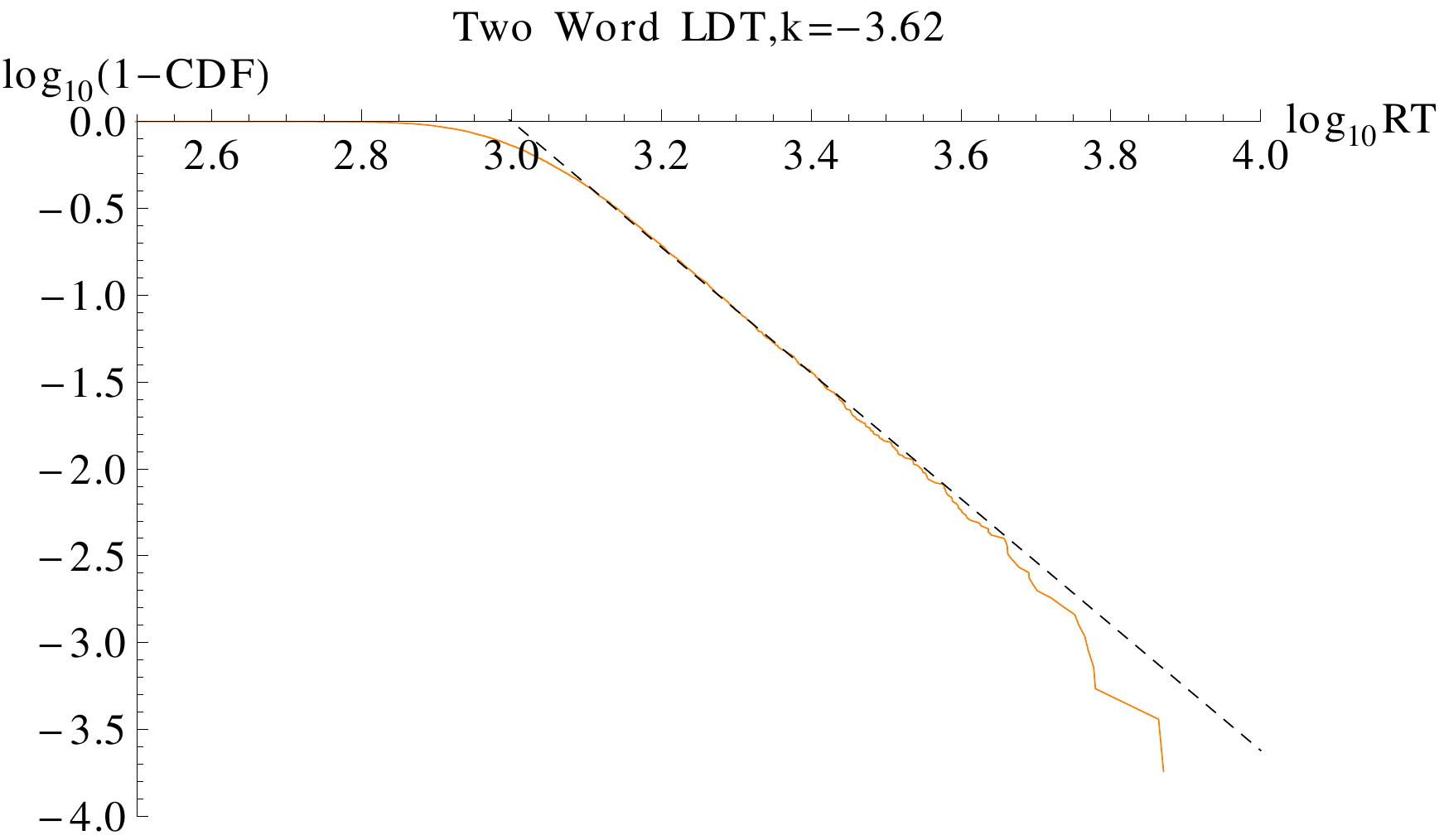}\\
\includegraphics[width=0.23\textwidth]{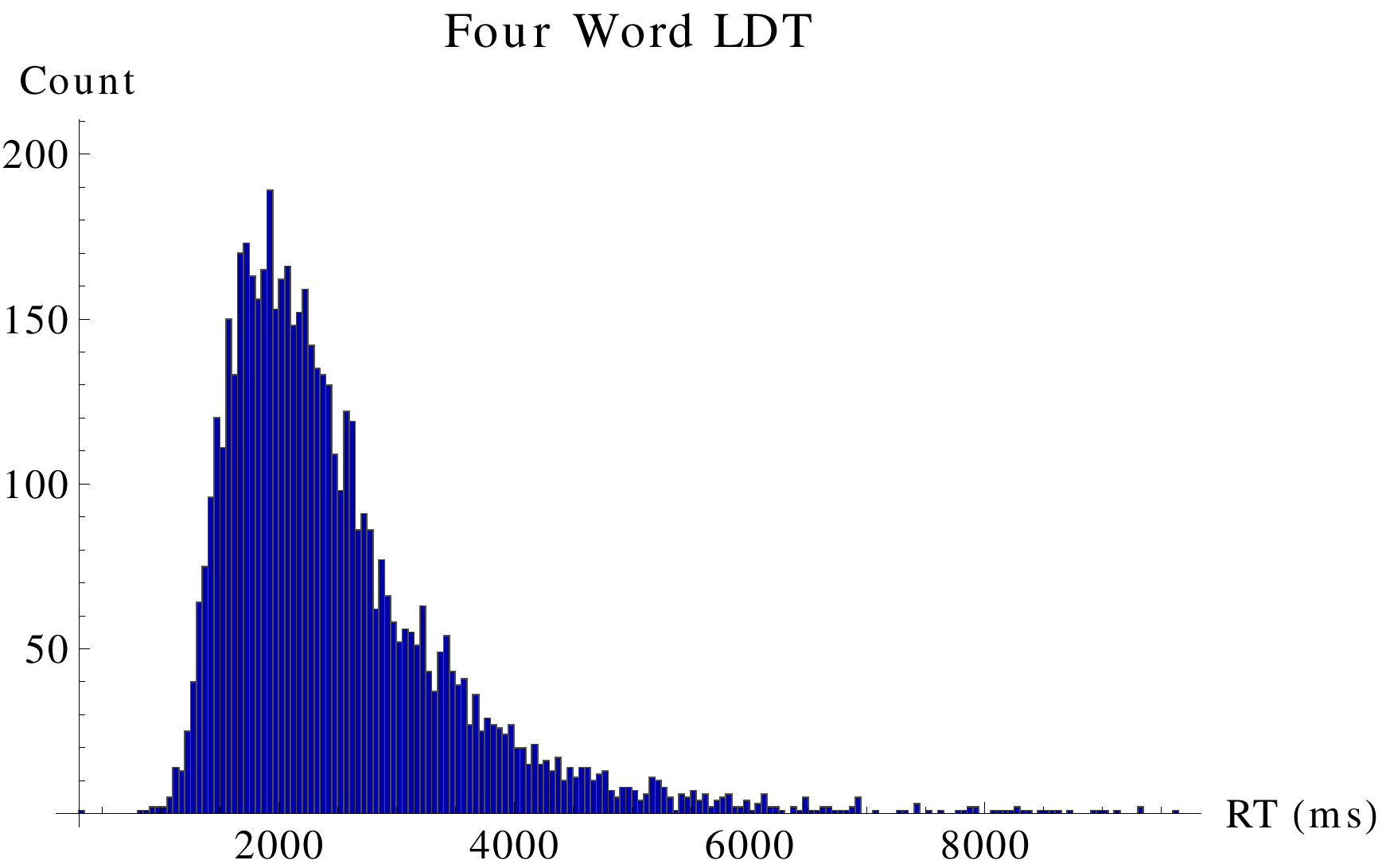}
\includegraphics[width=0.23\textwidth]{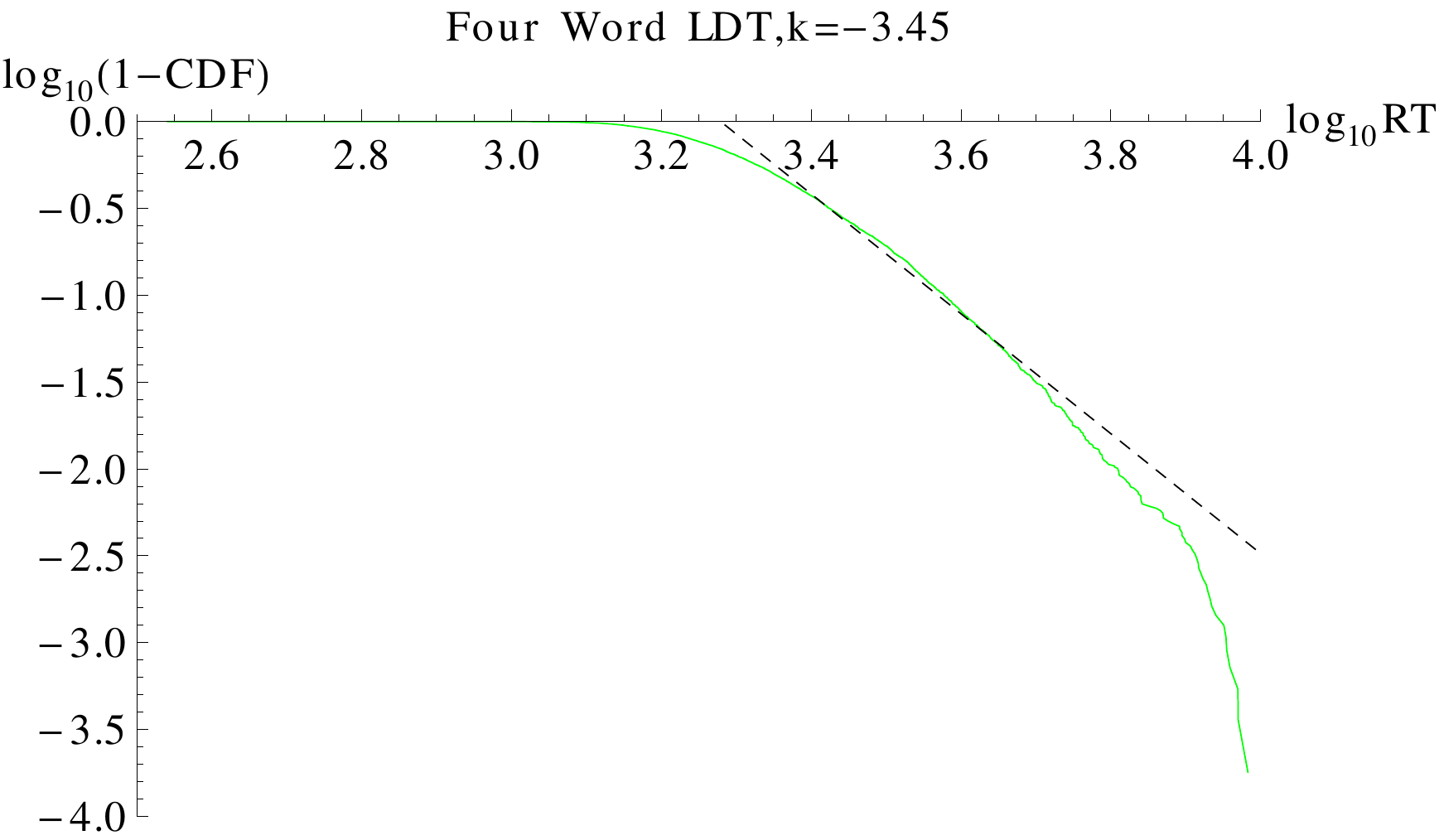}
\caption{Histogram and log-log plot of one, two, and four word LDT.}
\label{RT:fig:loglog_LDT}
\end{figure}

Log-log plot of power law tail fitting is discussed in Appendix \ref{Tail_Power}. 
In Figs. \ref{RT:fig:loglog_ELP}, \ref{RT:fig:loglog_Hick}, \ref{RT:fig:loglog_LDT}, we show the results for RT experiments. With the exception of LDT, trials for most of the tasks timed out by 4 or 5 seconds. This requirement has the potential to distort RT distributions, especially their slow tails, as log-log plot bends downward when RT is close to 4 seconds. (In the future, the requirement of maximum time limit should be dropped or, at least, the limiting time cutoff must increase to reflect the natural RT distribution.) In contrast, the maximum RT for LTD is approximately 10 seconds and, as seen In Fig. \ref{RT:fig:loglog_LDT} and Fig. \ref{RT:fig:GIGa_LDT}, the log-log plots are closer to straight lines and GIGa fit is good. 

\subsection{GIGa distribution fitting}

\begin{figure}[htp]
\centering
\includegraphics[width=0.45\textwidth]{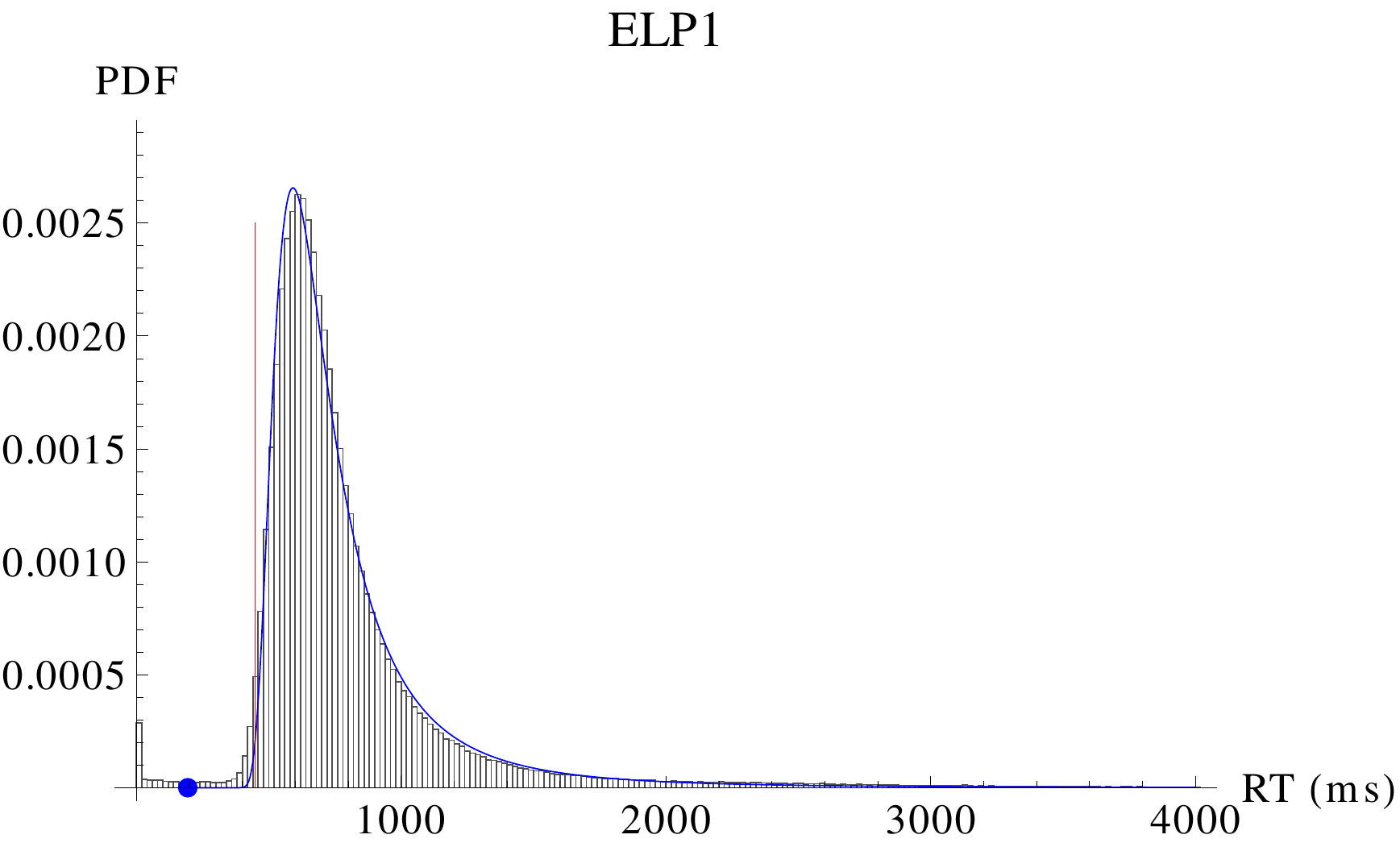}
\includegraphics[width=0.45\textwidth]{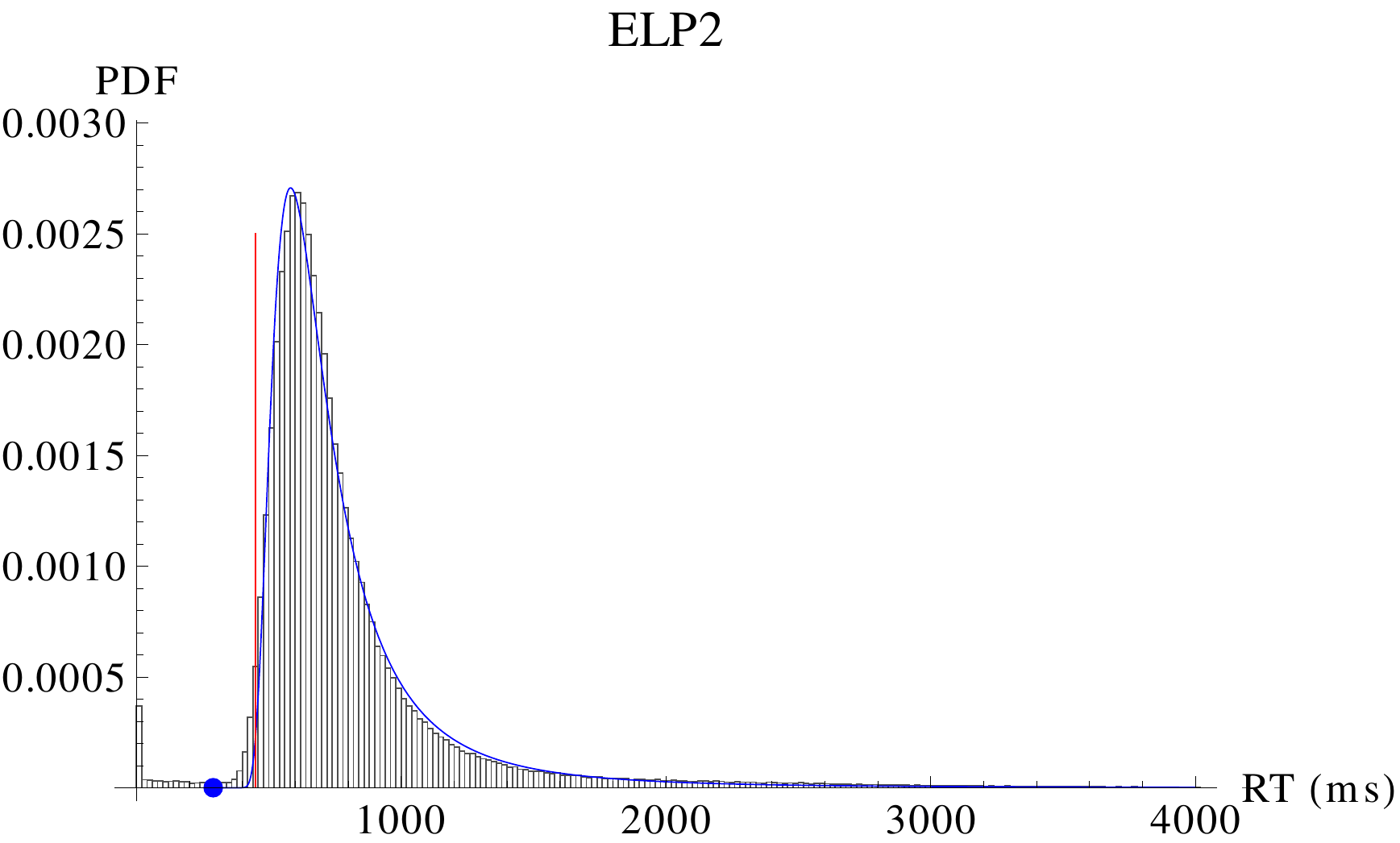}
\caption{GIGa fitting of ELP. ELP1: $\GIGa(0.73, 396, 3.69)$ with $\al\ga=2.7$. ELP2: $\GIGa(1.04, 345, 2.33)$ with $\al\ga=2.4$. The p-values are both 0.}
\label{RT:fig:GIGa_ELP}
\end{figure}

\begin{figure}[htp]
\centering
\includegraphics[width=0.45\textwidth]{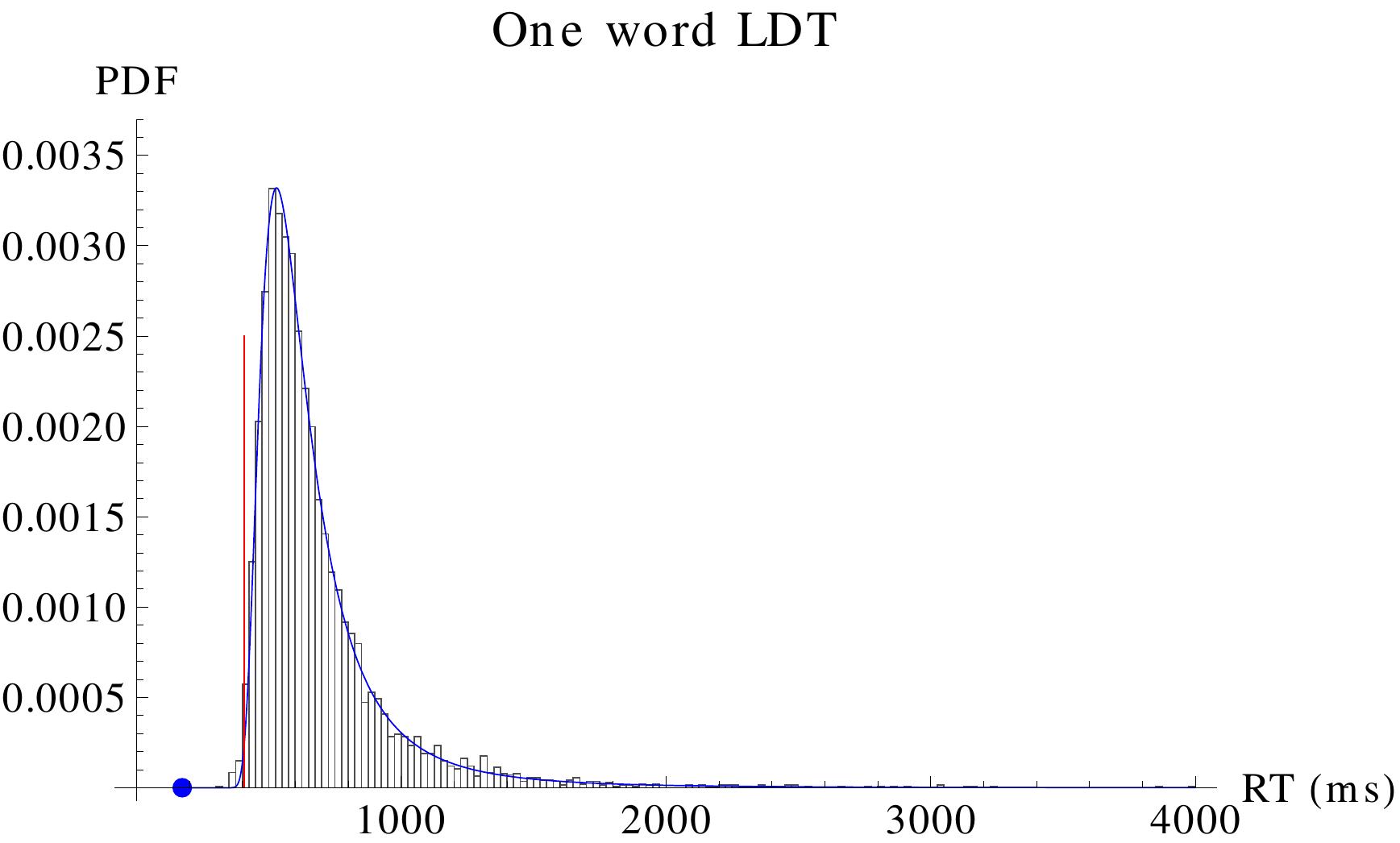}
\includegraphics[width=0.45\textwidth]{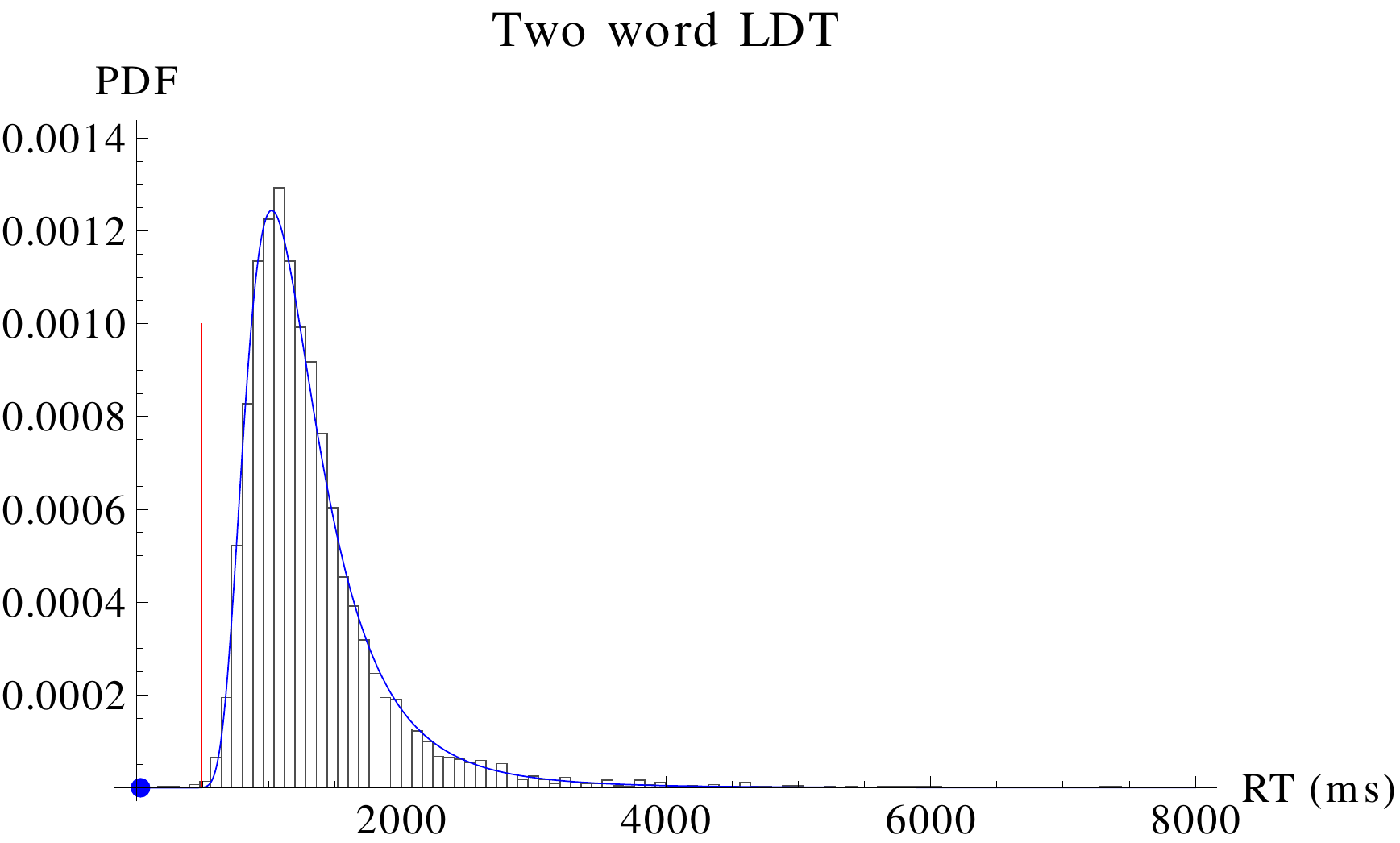}
\includegraphics[width=0.45\textwidth]{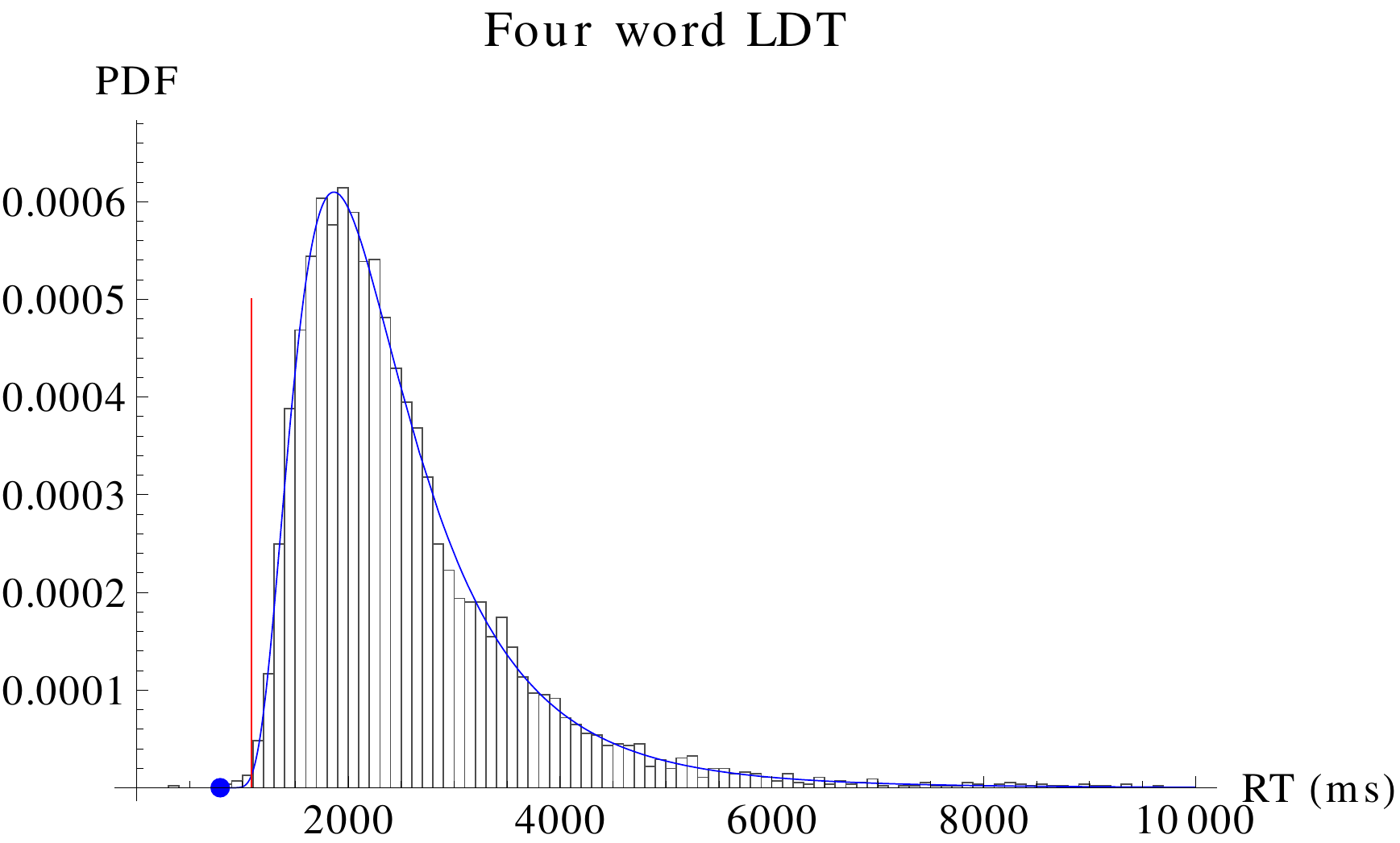}
\caption{GIGa fitting of one, two, four word LDT. One word LDT: $\GIGa(0.754, 357, 3.96)$ with $\al\ga=3.0$. Two word LDT: $\GIGa(1.96, 1424, 2.39)$ with $\al\ga=4.7$. Four word LDT: $\GIGa(25.1, 7.37\times 10^6, 0.376)$ with $\al\ga=9.4$. The p-values are 0.97, 0.82, and 0.87 respectively.}
\label{RT:fig:GIGa_LDT}
\end{figure}

\begin{figure}[htp]
\centering
\includegraphics[width=0.45\textwidth]{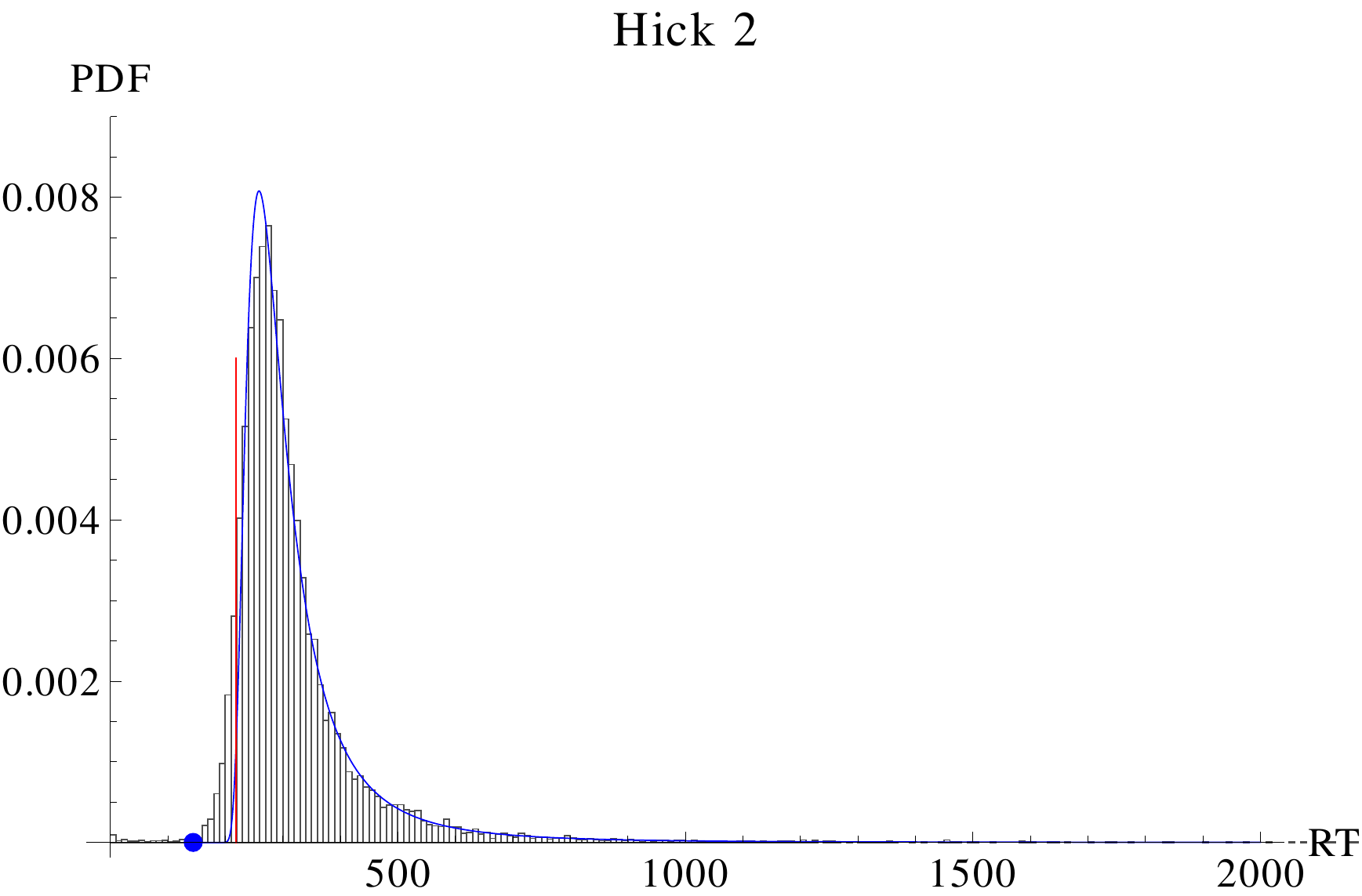}
\includegraphics[width=0.45\textwidth]{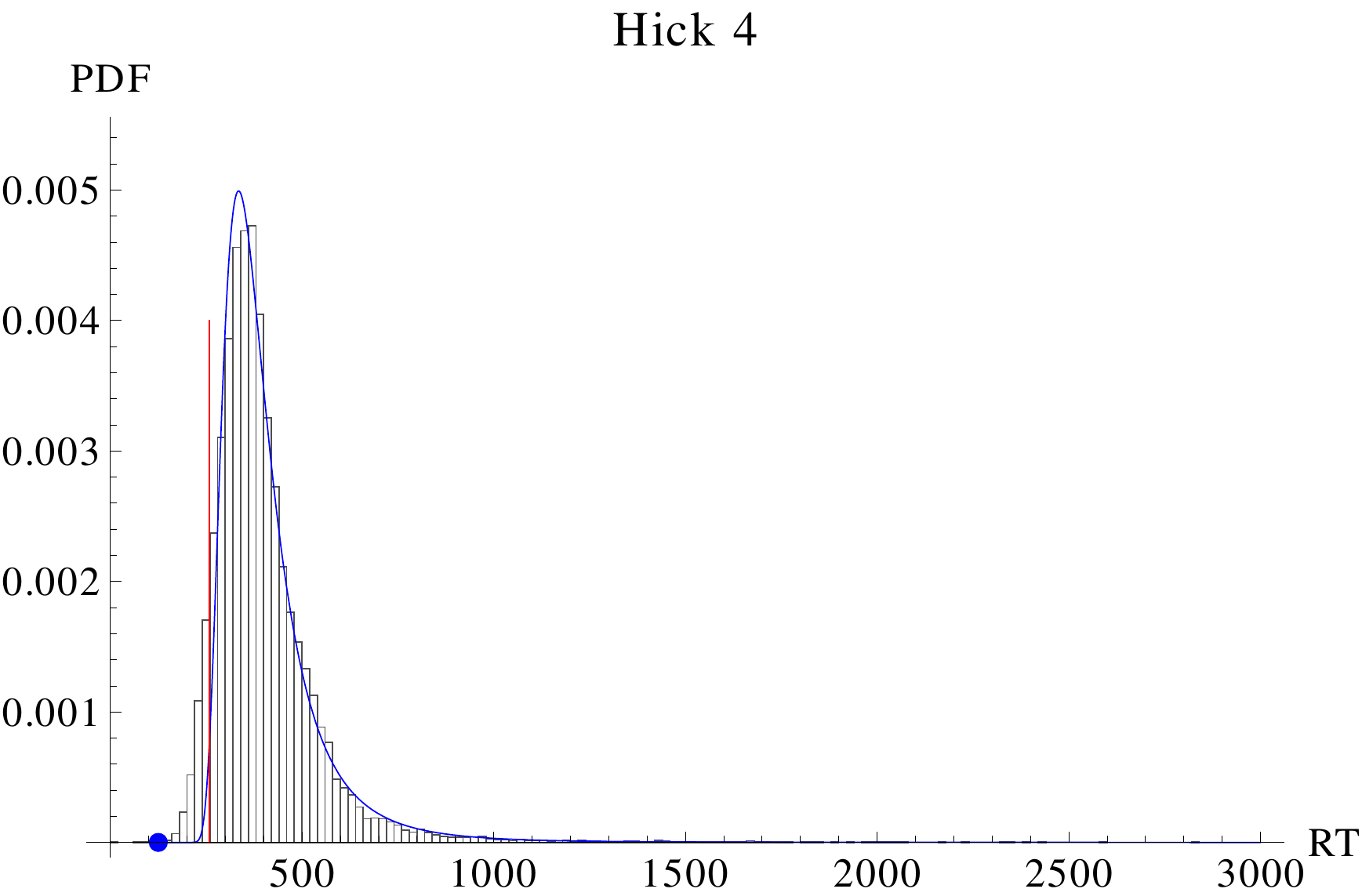}\\
\includegraphics[width=0.45\textwidth]{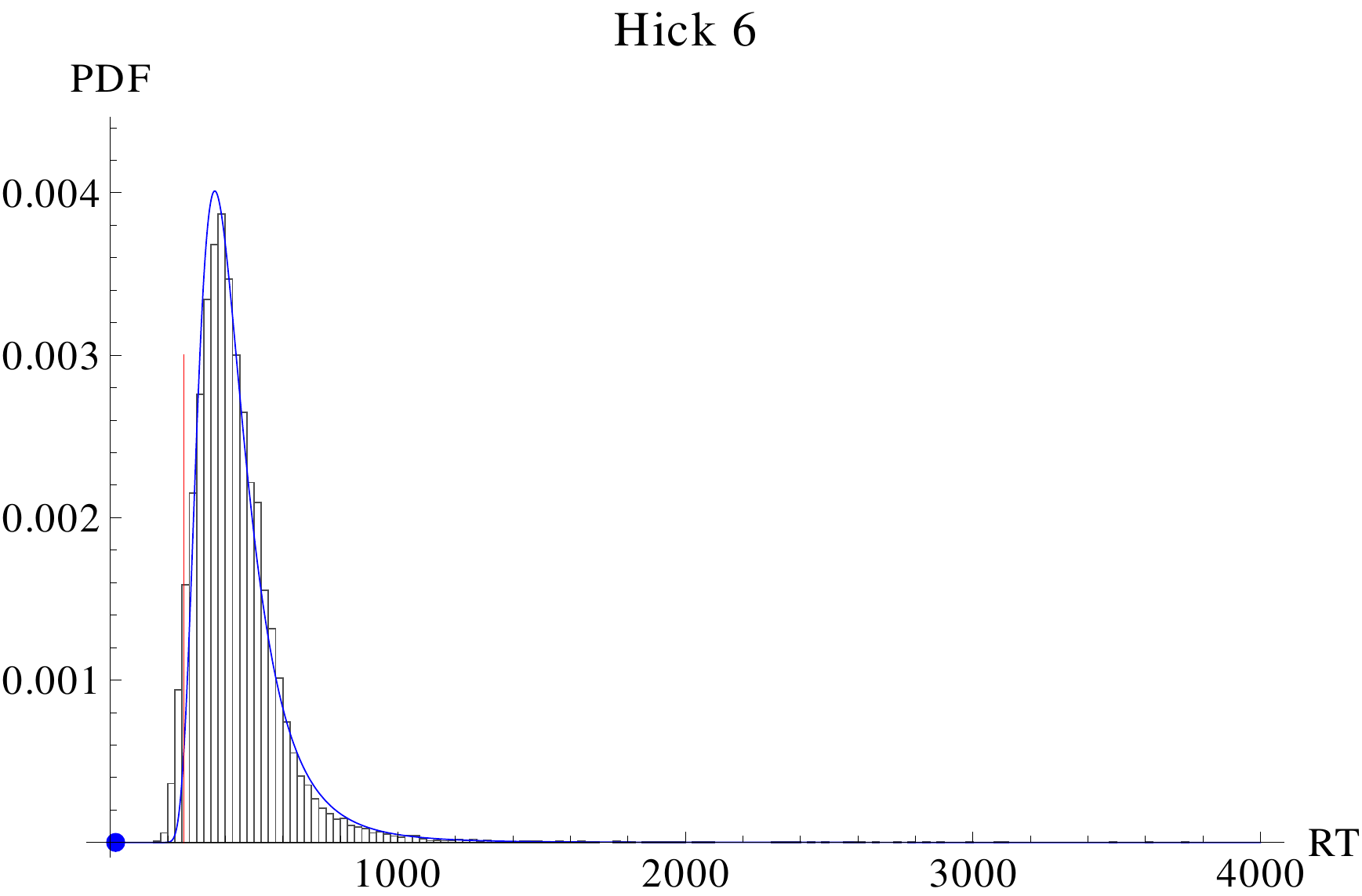}
\includegraphics[width=0.45\textwidth]{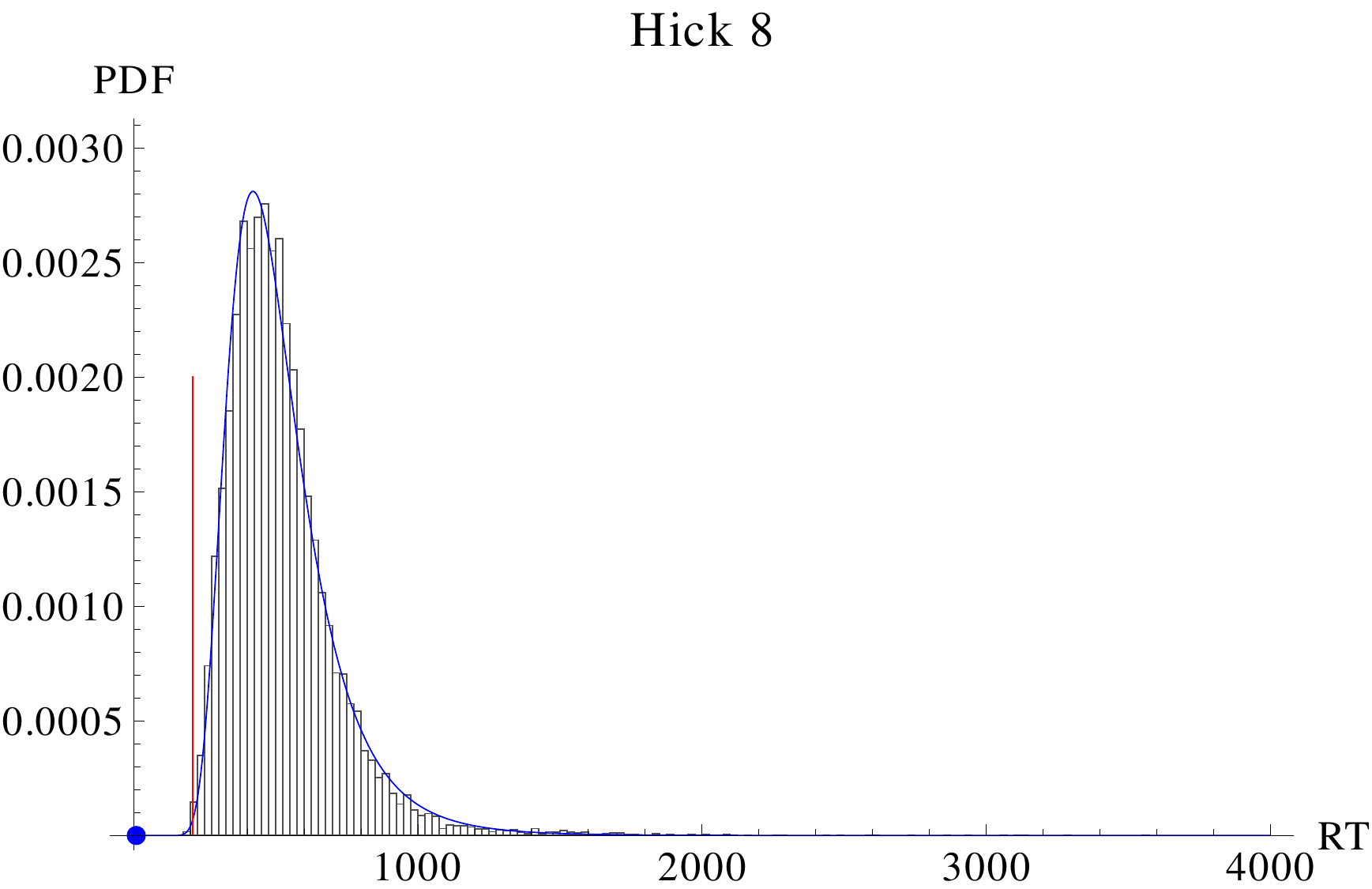}
\caption{GIGa fitting of Hick's experiment. The parameters $\{\al,\be,\ga\}$ of GIGa are 
$\{0.731, 115, 3.41\}$, $\{ 1.57, 275, 2.48 \}$, 
$\{ 1.64, 430, 3.07 \}$, and $\{ 7.80, 2922, 1.10\}$ respectively. 
$\al\ga$ are 2.5, 3.9, 5.0, and 8.6 respectively. The p-values are all 0.}
\label{RT:fig:GIGa_Hick}
\end{figure}

In Figs. \ref{RT:fig:GIGa_ELP}, \ref{RT:fig:GIGa_LDT}, and \ref{RT:fig:GIGa_Hick}, we show GIGa distribution  (Appendix \ref{GIGa_Scale}) fitting of RT. In the figures, the distance from the origin to the blue dot is rightward shift of GIGa distribution. The RTs to the left of the red lines are cut in the fitting of GIGa distribution. $\al,\be,\ga$, the cut and shift parameters are all found by minimizing the chi-squared test statistic as follows. We choose the cut and shift parameters, find $\al,\be,\ga$ through maximum likelihood estimation and compute the chi-squared test statistic. We repeat this process for another group of cut and shift parameters. In the end, we obtain the parameters that minimize the chi-squared test statistic.

\begin{figure}[htp]
\centering
\includegraphics[width=0.45\textwidth]{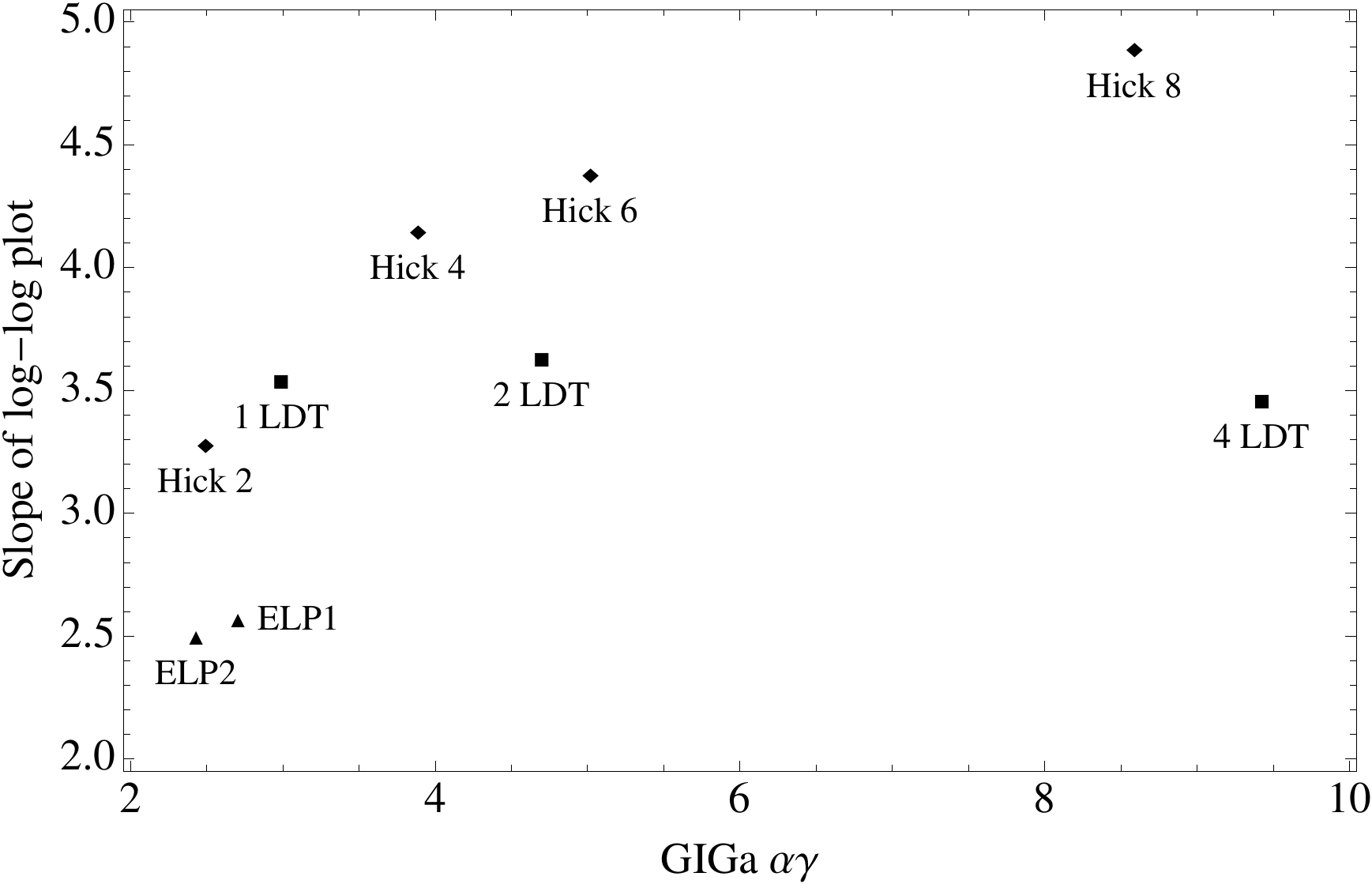}
\caption{Best fit GIGa $\al \ga$ versus log-log fitted tail exponent; triangles: ELP, squares: LDT, diamonds: HE}
\label{RT:alphagamma_loglog}
\end{figure}

Visually, GIGa fitting is good, yet p-values are all zero with the exception of LDT. As discussed above, a possible explanation is that the participants are not given enough time to respond, which distorts RT distributions. Also, Ref. \cite{vanzandt2000} argues that chi-squared statistic yields poor results for goodness-of-fit -- we used chi-squared statistic because, due to the cut parameter, the total number of RTs is not fixed in our parameter fitting. Lastly, in Fig. \ref{RT:alphagamma_loglog} we show the the relationship between the tail exponent parameter $\al \ga$ and log-log fitted exponent parameter -- with the exception of 4 LDT (which is one of the hardest tasks -- see below), the correspondence is quite good.

\section{Task difficulty}

\begin{figure}[htp]
\centering
\includegraphics[width=0.45\textwidth]{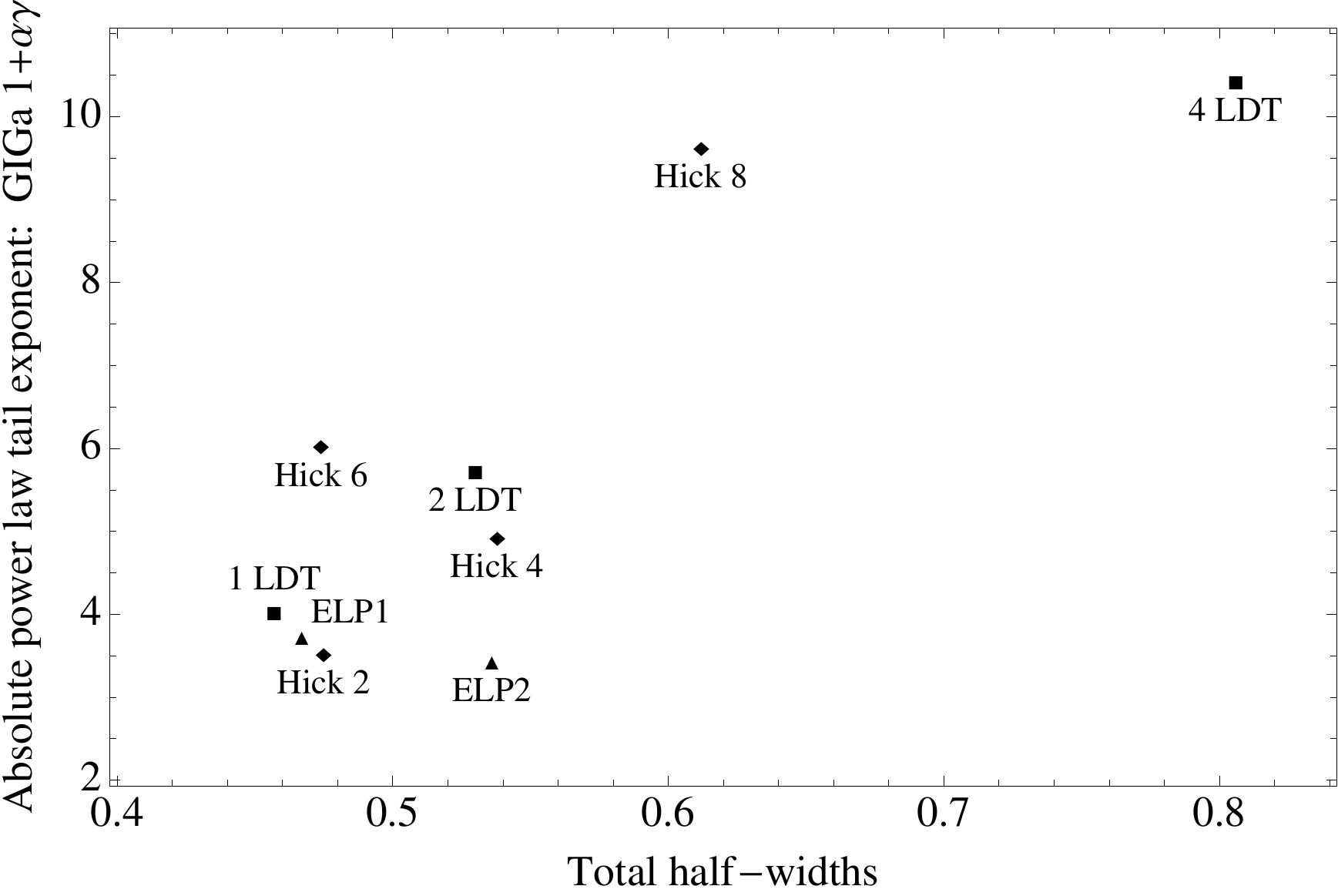}
\caption{Best fit GIGa absolute power law tail exponent $\al \ga +1$ versus its half width; triangles: ELP, squares: LDT, diamonds: HE}
\label{RT:half_width_alphagamma}
\end{figure}

In Fig. \ref{RT:half_width_alphagamma}, we plot the power law exponent from the best fit GIGa above as a function of their half-width. With the exception of Hick 6, there is a clear tracking between the two (notice that by eye HE PDFs seemingly show decrease of modal PDF and increase of PDF half-width with the increase of Hick's number).

We speculate that the half width of the distribution would be a natural measure of a task difficulty. This is easily analyzed in terms of the GIGa distribution, which we believe is well suited to description of RT distributions. In Appendix \ref{GIGa_Scale}, it is explained that due to GIGa's scaling property, it is sufficient to consider the $\ga=1$ case, that is IGa. Furthermore, we can eliminate one more parameter by setting mean to unity. In some cognitive tasks, the mean may not be a good indicator of difficulty since an easy cognitive task may require a more idiosyncratic response and vice versa.

For such IGa, a single parameter $\al$ then defines both scale and shape, that is the half width is directly relates to the exponent of the power law tail. As seen in Fig. \ref{IGa_halfwidth} it has a maximum as a function of this parameter, which also marks a crossover between IGa limiting behaviors. This opens up an interesting possibility that depending on the magnitude of $\al$, increase in the task difficulty may either increase or decreases the magnitude of the power law exponent. This subject, including sufficient data to analyze the aforementioned scaling property, requires further investigation.\\

\section{Acknowledgments}
J.G. Holden's work was supported by the National Science Foundation Award BCS-0642718.\\ 

{\bfseries{We repeat a number of Appendices from \cite{ma13ST} verbatim, given that the expected audiences for these two papers are vastly different.} }

\appendix

\section{Properties of GIGa distribution}\label{GIGa_Scale}
\begin{figure}[htp]
\centering
\includegraphics[width=0.23\textwidth]{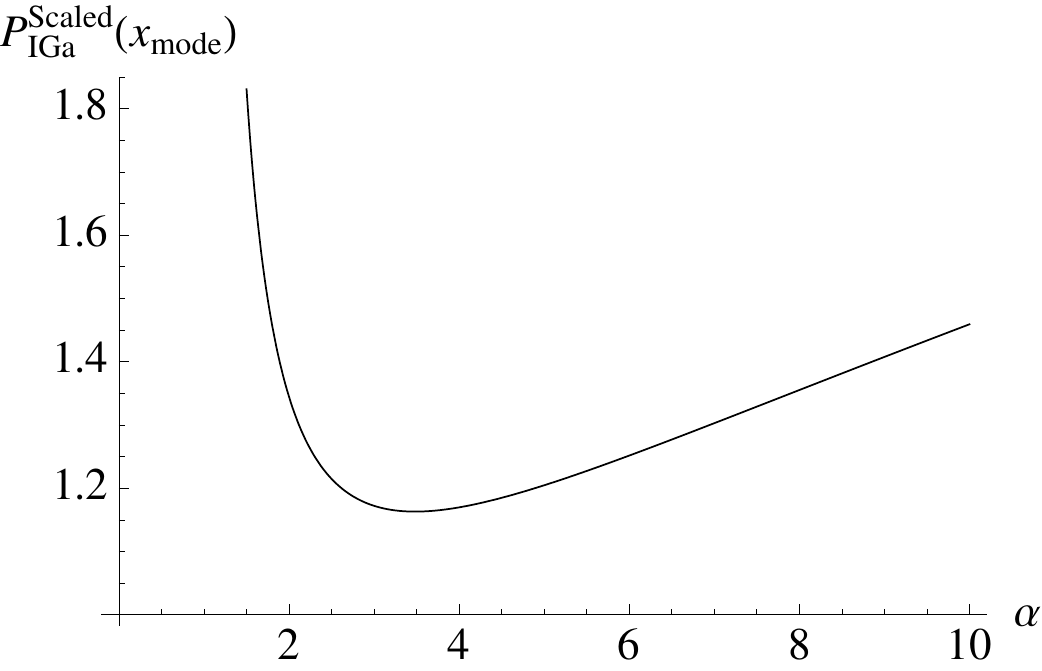}
\includegraphics[width=0.23\textwidth]{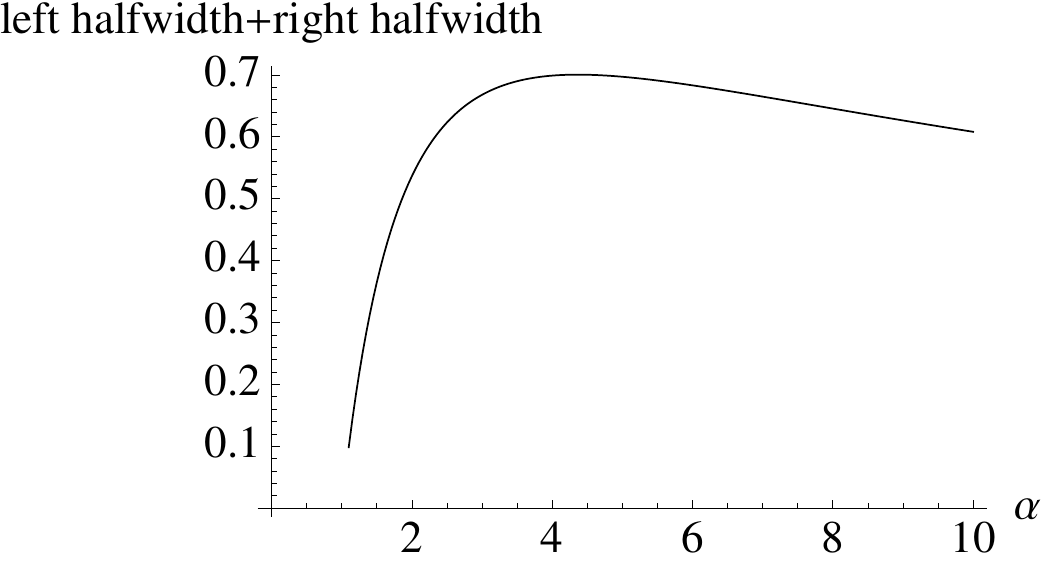}
\caption{Mode and half-width of scaled IGA as a function of $\al$}\label{ST:fig:IGa_PDF_mode}
\label{IGa_halfwidth}
\end{figure}

\begin{figure}[htp]
\centering
\includegraphics[width=0.345\textwidth]{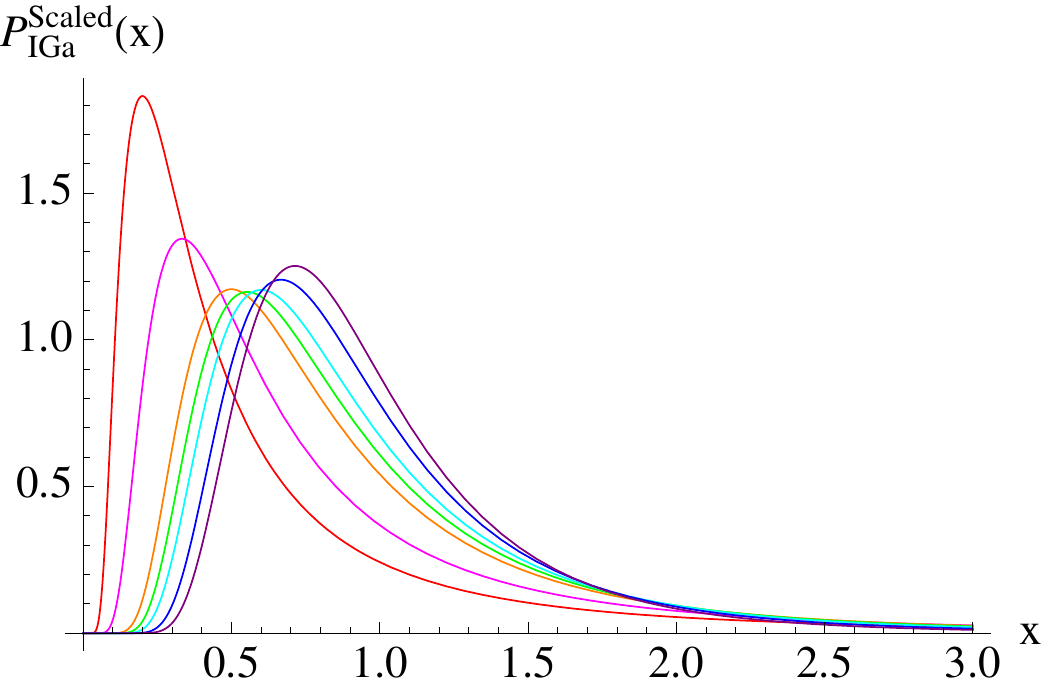}
\caption{PDF of IGa distributions. From left to right, $\al=1.5, 2, 3, 3.48, 4, 5$, and $6$, corresponding to red, magenta, orange, green, cyan, blue, and purple lines.}
\label{ST:fig:IGa_PDF_list}
\end{figure}

We begin with the $\ga=1$ limit of GIGa, namely IGa distribution PDF
\begin{equation}
P_{\IGa}(x) = \fr{1}{\be \Gamma(\al)} \exp\lf[ -\fr{\be}{x} \rg] \lf(\fr{\be}{x}\rg)^{1+\al} .
\end{equation}
Setting the mean to unity, the scaled distribution is 
\begin{equation}
P_{\IGa}^{\text{Scaled}}(x) = \fr{(\al-1)^\al \exp\lf( -\fr{\al-1}{x} \rg)}{\Gamma(\al) x^{1+\al}} . 
\end{equation}
The mode of the above distribution is $x_{\text{mode}} = (\al-1)/(\al+1)$. The modal PDF is
\begin{equation}
P_{\IGa}^{\text{Scaled}}(x_\text{mode}) = 
\fr{(1+\al)^{1+\al} \exp(-1-\al)}{\Gamma(\al)(\al-1)} , 
\end{equation}
which has a minimum at $\al \approx 3.48$ as shown in Fig. \ref{ST:fig:IGa_PDF_mode}. The change in PDF behavior on transition through this value is clearly observed in Fig. \ref{ST:fig:IGa_PDF_list}. Also plotted in Fig. \ref{ST:fig:IGa_PDF_mode} is the half-width of the distribution. Clearly, it highly correlates with the PDF maximum above.

Both minimum and maximum above clearly separate the regime of small $\al$: $\al\rightarrow 1$, where the approximate form of the scaled PDF is 

\begin{equation}
P_{\IGa}^{\text{Scaled}}(x) \approx
\fr{(\al-1) \exp\lf[ -\fr{\al-1}{x} \rg]}{x^2}
\end{equation}
whose mode is $(\al-1)/2$ and the magnitude of the maximum is $ 4 \exp[-(\al-1)^2/2]/(\al-1) \approx 4/(\al-1) $, from the regime of large $\al$, $\al\rightarrow \infty$, where 
\begin{equation}
P_{\IGa}^{\text{Scaled}}(x) \rightarrow \delta(x-1) .
\end{equation}

\begin{figure}[htp]
\centering
\includegraphics[width=0.345\textwidth]{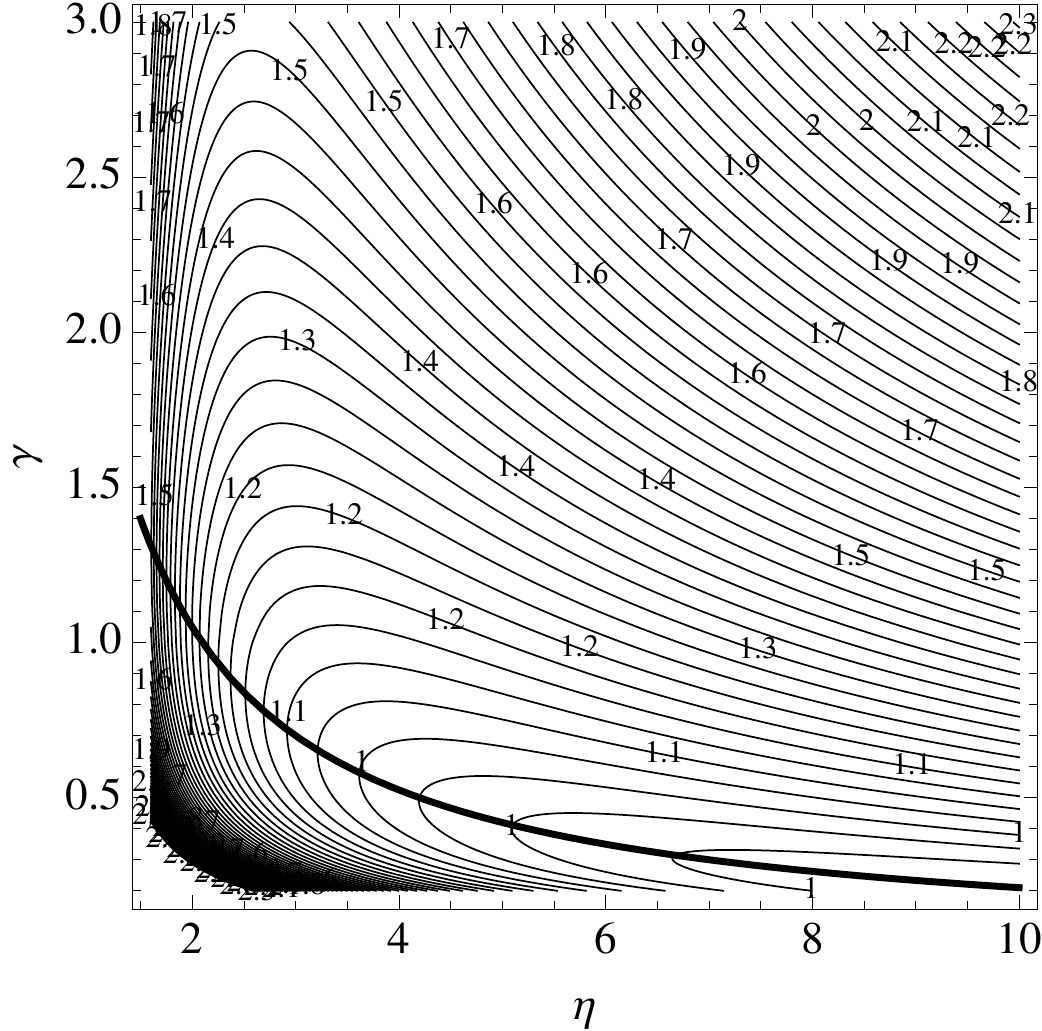}
\includegraphics[width=0.345\textwidth]{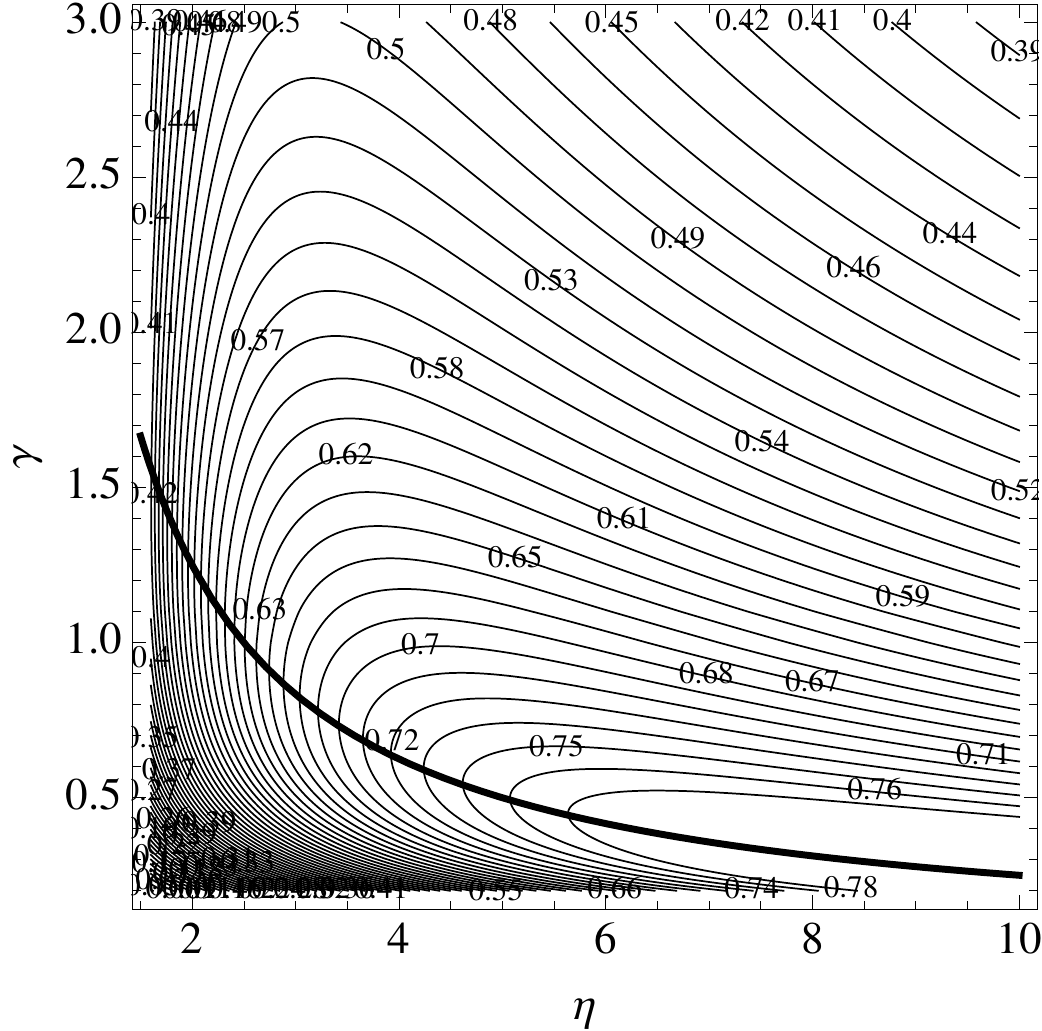}
\caption{Top: contours of modal PDF of GIGa distributions with mean $1$. Thin lines: contours of modal PDF. Thick line: $\ga=2.1/\eta$. Bottom: contours of total half-widths of GIGa distributions with mean $1$. Thick line: $\ga=2.5/\eta$. }
\label{ST:fig:GIGa_contour}
\end{figure}

\begin{figure}[htp]
\centering
\includegraphics[width=0.345\textwidth]{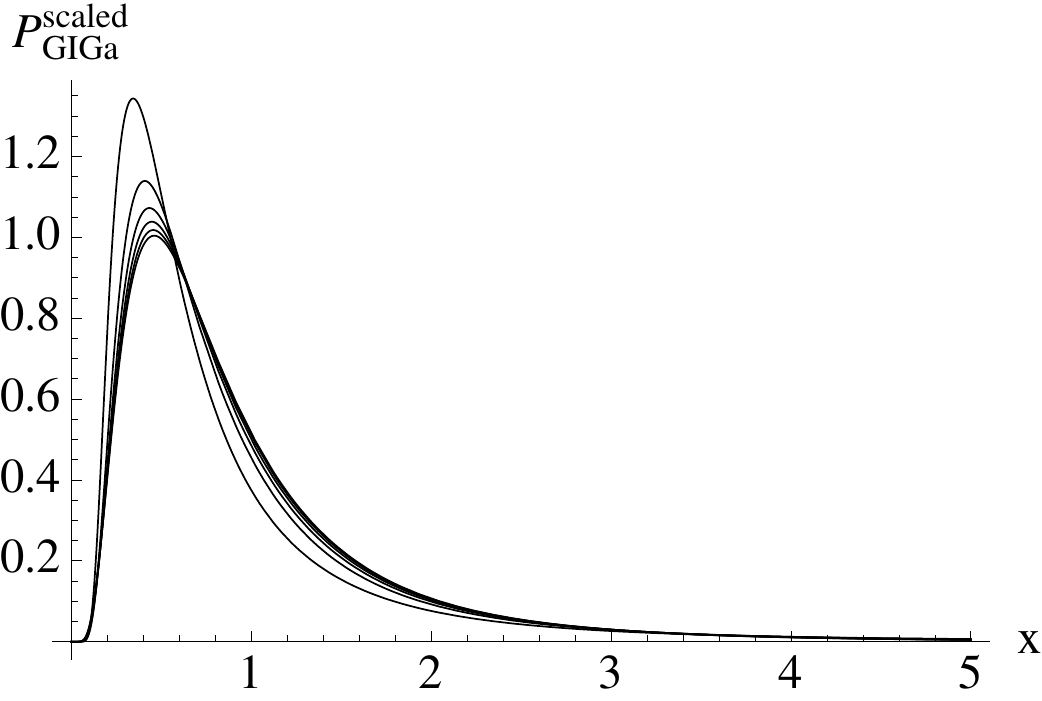}
\caption{Scaled PDF of GIGa distributions with mean $1$. In the plots, $\ga=2.1/\eta$. Six lines correspond to $\eta=2,3,...,7$}
\label{ST:fig:GIGa_PDF_list_overlay}
\end{figure}

\begin{figure}
\centering
\includegraphics[width=0.23\textwidth]{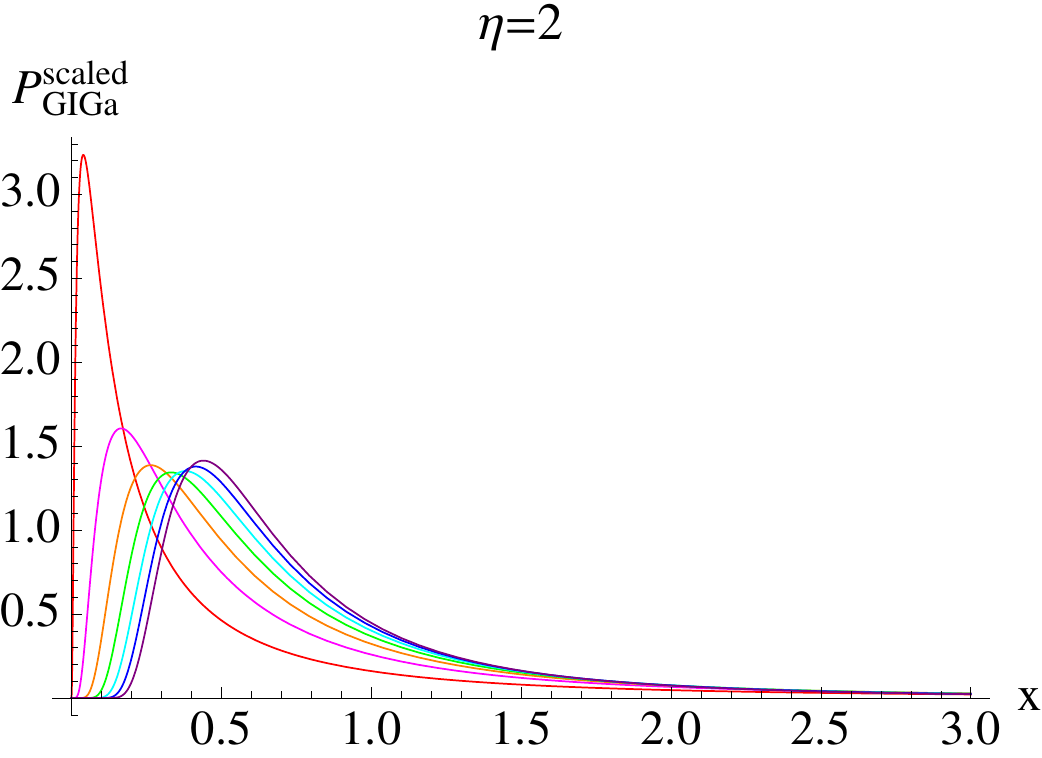}
\includegraphics[width=0.23\textwidth]{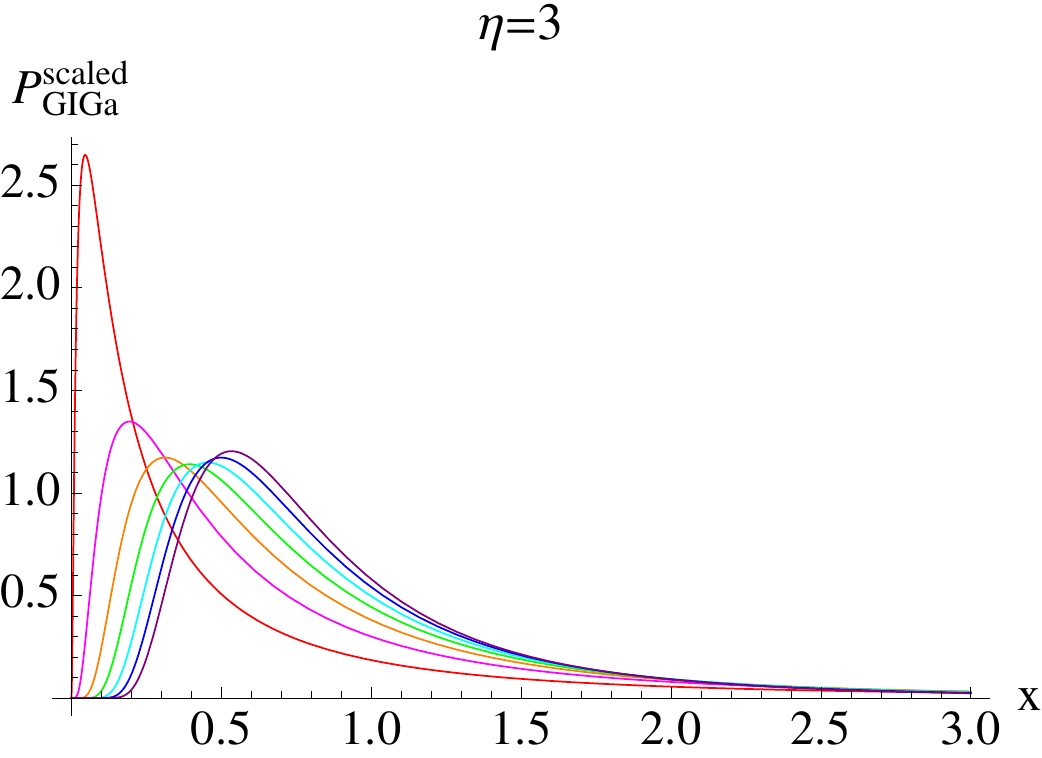}\\
\includegraphics[width=0.23\textwidth]{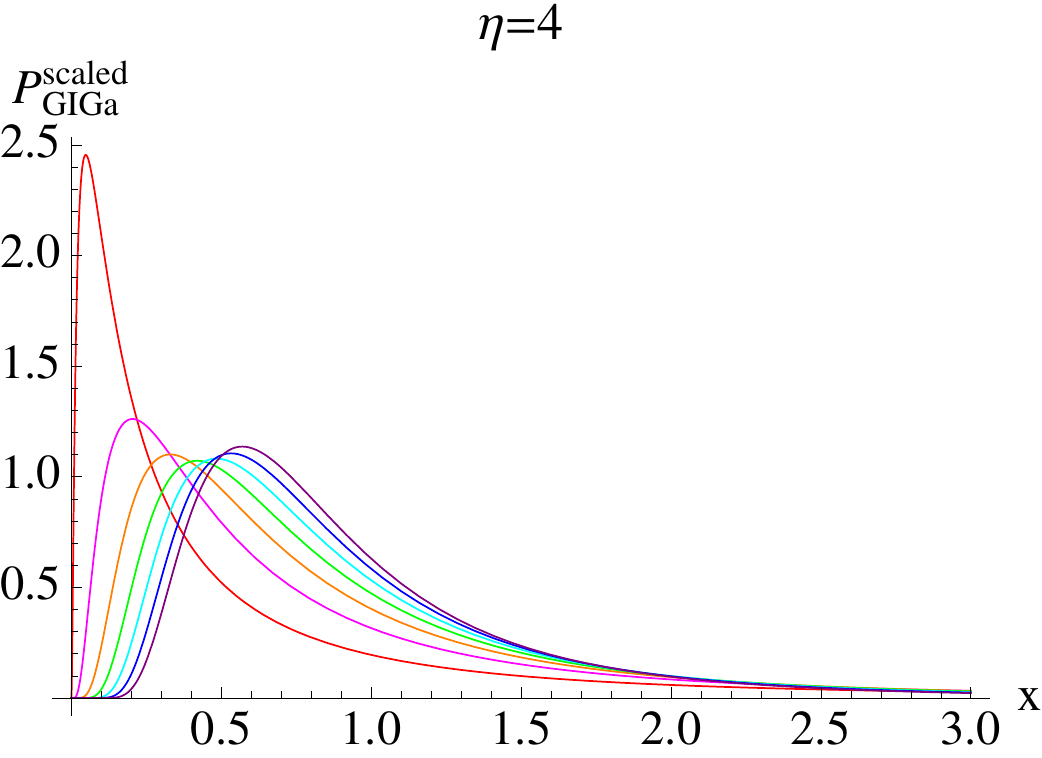}
\includegraphics[width=0.23\textwidth]{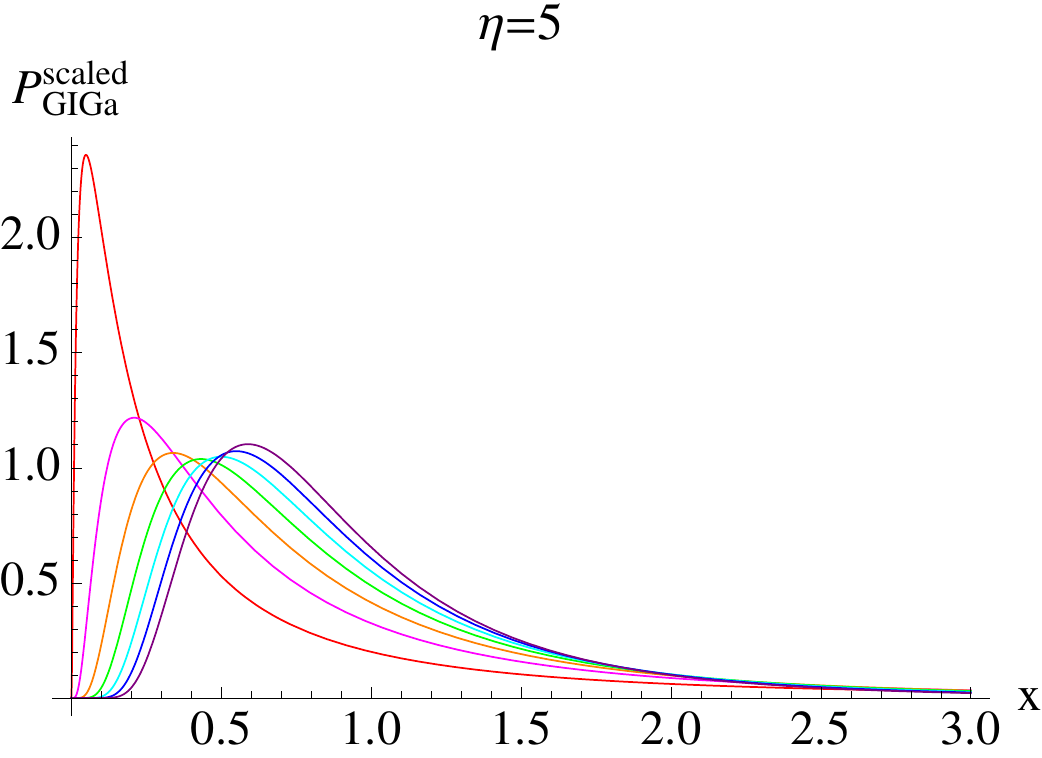}\\
\includegraphics[width=0.23\textwidth]{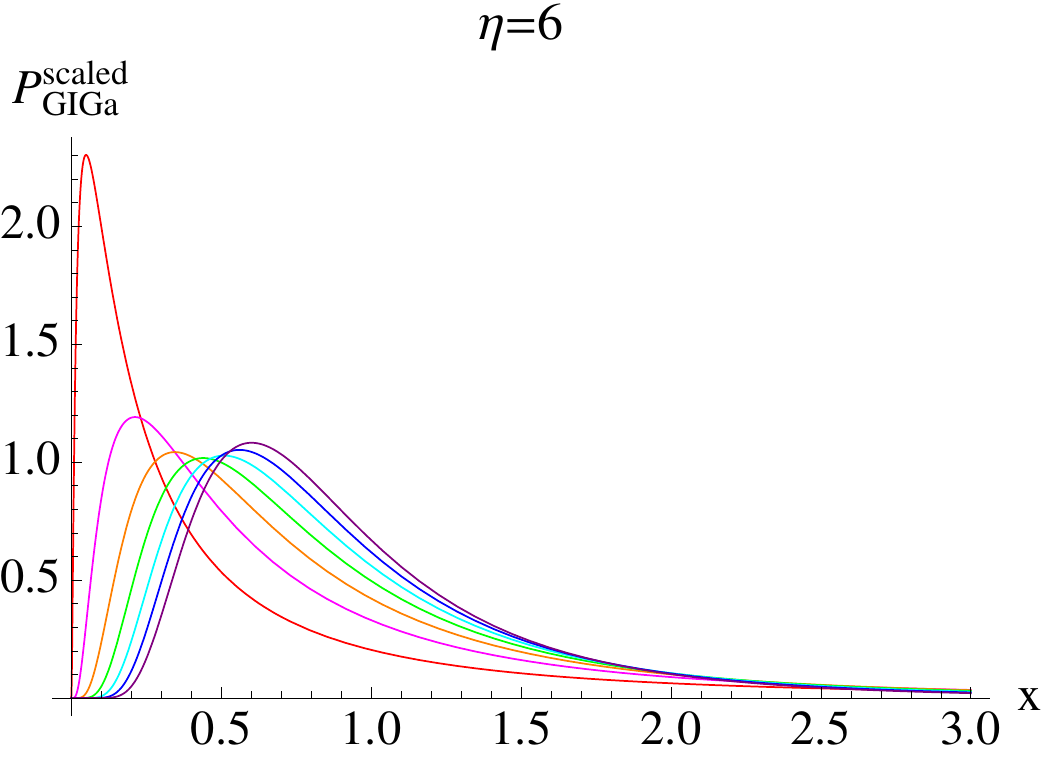}
\includegraphics[width=0.23\textwidth]{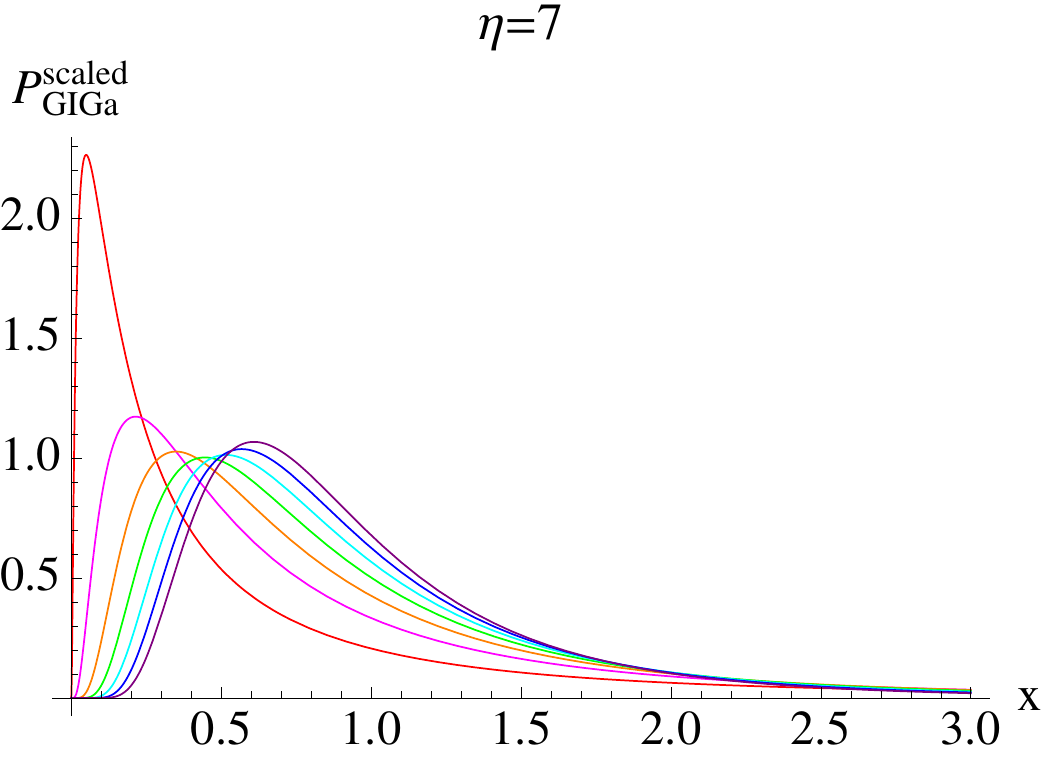}
\caption{Scaled PDF of GIGa distributions with mean $1$. In each subplot with constant $\eta$, from left to right, $\ga=0.5/\eta, 1/\eta, 1.5/\eta, 2/\eta, 2.5/\eta, 3/\eta$, and $3.5/\eta$, corresponding to red, magenta, orange, green, cyan, blue, and purple lines. }
\label{ST:fig:GIGa_PDF_list}
\end{figure}

We now turn to GIGa distribution and the effect of parameter $\ga$. In Fig. \ref{ST:fig:GIGa_contour} we give the contour plots of modal PDF and total half-widths in the $(\eta, \ga)$ plane, where $\eta = \al\ga$ and $-1-\eta$ is the exponent of the power law tail.  We observe an interesting \emph{scaling property} of GIGa: for $\ga\approx 2.1/\eta$, the dependence of the PDF on $\eta$ is very weak, as demonstrated in Fig. \ref {ST:fig:GIGa_PDF_list_overlay}, where it is plotted for integer $\eta$ from 2 to 7. An alternative way to illustrate this is to plot PDF for a fixed $\eta$ and variable $\ga$, as shown in Fig. \ref{ST:fig:GIGa_PDF_list}. Following the thick line we notice that, for $\eta > 3$, mode and half-width change very little with $\eta$. The key implication of the scaling property is that IGa contains all essential features pertinent to GIGa.

\section{Parametrization of the GIGa family of distributions}\label{GIGa_LN}

This Appendix is a self-contained re-derivation of a LN limit of GIGa.  \cite{weibullcom} The three-parameter GIGa distribution is given by
\begin{equation}\label{ST:eq:PDF_GIGa}
\GIGa(x; \al, \be, \ga)
= \fr{\ga}{\be\Ga(\al )} e^{-\lf(\fr{\be }{x}\rg)^\ga} \lf(\fr{\be }{x}\rg)^{1+\al\ga}
\end{equation}
for $x>0$ and 0 otherwise.
We require that $\al, \be, \ga >0$. IGa is the the $\ga=1$ case of GIGa:
\begin{equation}\label{ST:eq:PDF_IGa}
\IGa(x; \al, \be)
= \fr{1}{\be\Ga(\al )} e^{-\fr{\be }{x}} \lf(\fr{\be }{x}\rg)^{1+\al} .
\end{equation}
Note that GIGa and IGa have power-law tails $x^{-1-\al\ga}$ and $x^{-1-\al}$ respectively for $x \gg \be$.

We proceed to rewrite GIGa in the following form:
\begin{equation}
\begin{split}
\GIGa(x; \al, \be, \ga)
&= \fr{\ga}{x \Ga(\al )} \exp\lf[\al \ln\lf(\fr{x}{\be}\rg)^{-\ga} - \lf(\fr{x}{\be}\rg)^{-\ga}\rg] .
\end{split}
\end{equation}
A re-parameterization
\begin{eqnarray}\label{ST:eq:reparameterization}
\mu &=& \ln \be - \fr{1}{\ga} \ln\fr{1}{\lambda^2}  \\
\si &=& \fr{1}{\ga \sqrt{\al}} \\
\lambda &=& \fr{1}{\sqrt{\al}} ,
\end{eqnarray}
with $\si>0$ and $\lambda>0$, allows to express the old parameters in terms of the new:
\begin{eqnarray}
\al &=& \fr{1}{\lambda^2} \\
\be &=& e^\mu \lambda^{-\fr{2\si}{\lambda}} \\
\ga &=& \fr{\lambda}{\si}
\end{eqnarray}
leading, in turn, to
\begin{equation}\label{ST:eq:exp_expansion}
\lf(\fr{x}{\be}\rg)^{-\ga} = e^{-\fr{\lambda}{\si} (\ln x - \mu)} \lambda^{-2}
\end{equation}
\begin{equation}
\ln\lf(\fr{x}{\be}\rg)^{-\ga} = -\fr{\lambda}{\si} (\ln x - \mu) + \ln(\lambda^{-2}) \\
\end{equation}
and 
\begin{equation}\label{ST:eq:GIGaLN1}
\al \ln\lf(\fr{x}{\be}\rg)^{-\ga} - \lf(\fr{x}{\be}\rg)^{-\ga} \\
\approx \fr{\ln(\lambda^{-2}) - 1}{\lambda^2} - \fr{(\ln x - \mu)^2}{2\si^2} ,\\
\end{equation}
where we have used the Taylor expansion of the $\exp$ term in Eq. (\ref{ST:eq:exp_expansion}), which depends on $\lambda/\si = \ga \rightarrow 0^+.$
We can also prove that
\begin{equation}\label{ST:eq:GIGaLN2}
\fr{\ga}{\Ga(\al)} \exp\lf[ \fr{\ln(\lambda^{-2}) - 1}{\lambda^2} \rg] = \fr{1}{\sqrt{2\pi}\si} ,
\end{equation}
based on the Stirling's approximation when we let $\lambda^{-2} = \al \rightarrow +\infty$.

Upon substitution of Eqs. (\ref{ST:eq:GIGaLN1}) and (\ref{ST:eq:GIGaLN2}) into Eq. (\ref{ST:eq:PDF_GIGa}), we obtain the LN distribution
\begin{equation}\label{ST:eq:LN}
\LN(x; \mu, \si) = \fr{1}{\sqrt{2\pi}\si x} \exp \lf[ - \fr{(\ln x - \mu)^2}{2 \si^2} \rg] .
\end{equation}
In conclusion, GIGa has the limit of LN when $\lambda$ tends to $0$ in such a way that $\alpha$ tends to $+\infty$ quadratically and $\gamma$ tends to $0$ linearly.

GIGa (IGa) are also transparently related to GGa (Ga) distribution: $\GGa(x; \al, \be, \ga) \xleftrightarrow{\ga\leftrightarrow -\ga} -\GGa(x; \al, \be, -\ga) = -\GIGa(x; \al, \be, \ga)$ and $\GGa(x; \al, \be, \ga) \leftrightarrow \GIGa(1/{x}; \al, 1/{\be}, \ga)$. Note, finally, that Lawless \cite{Lawless1982} derived the LN limit of GGa in a manner similar to ours, which solidifies the concept of the ``family'' that unites these distributions.

\section{Stochastic ``birth-death'' model }\label{Birth_Death}
Many natural and social phenomena fall into a stochastic ``birth-death'' model, described by the equation 
\begin{equation}
dx = c_1 x^{1-\ga} dt - c_2 x dt + \si x dW
\end{equation}
where $x$ can alternatively stand for such additive quantities as wealth, \cite{ma2013distribution} mass of a species, \cite{west2012} volatility variance, \cite{ma13ST} etc., and cognitive response times here.

The second term in the rhs describes an exponentially fast decay, such as the loss of wealth and mass due to the use of one's own resources, or the reduction of volatility in the absence of competing inputs and of response times due to learning. The first rhs term may alternatively describe metabolic consumption, acquisition of wealth in economic exchange, plethora of market signals, and variability of cognitive inputs. 

The third, stochastic term is the one that changes the otherwise deterministic approach, characterized by the saturation to a final value of the quantity, with the probabilistic distribution of the values - as it were, GIGa in the steady-state limit. Furthermore, just as the wealth model has microscopic underpinnings in a network model of economic exchange, \cite{ma2013distribution} it is likely that stochastic ontogenetic mass growth \cite{west2012} could be described by analogous network model based on capillary exchange. A network analogy may be possible for cognitive response times and volatility as well.

\section{Log-log plot of distribution tails}\label{Tail_Power}

\begin{figure}[htp]
\centering
\includegraphics[width=0.345\textwidth]{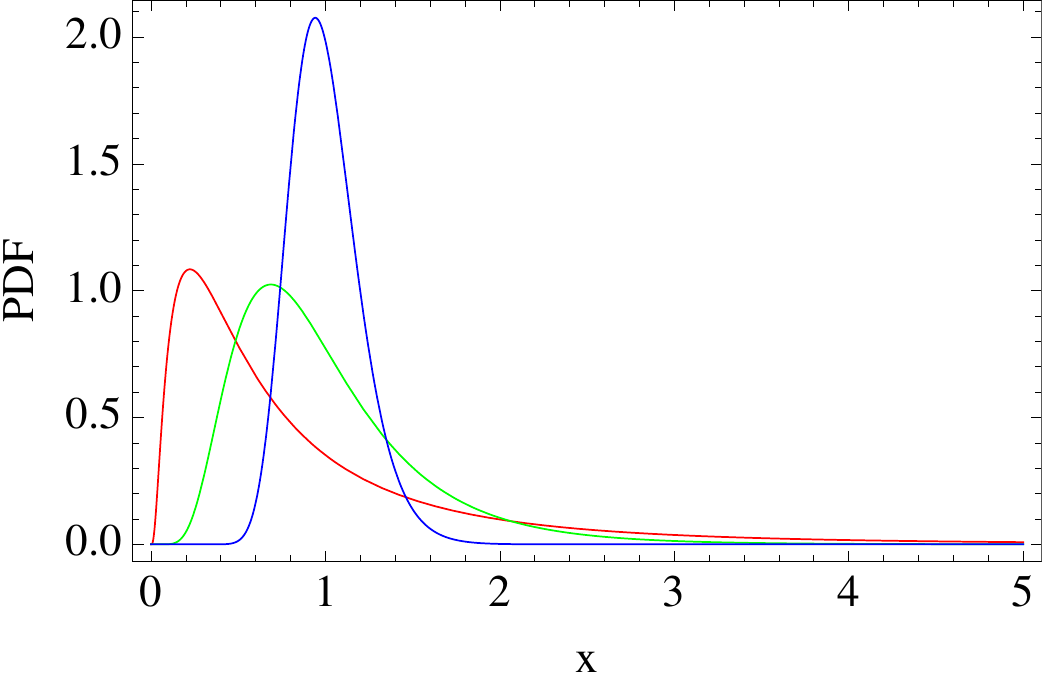}
\includegraphics[width=0.345\textwidth]{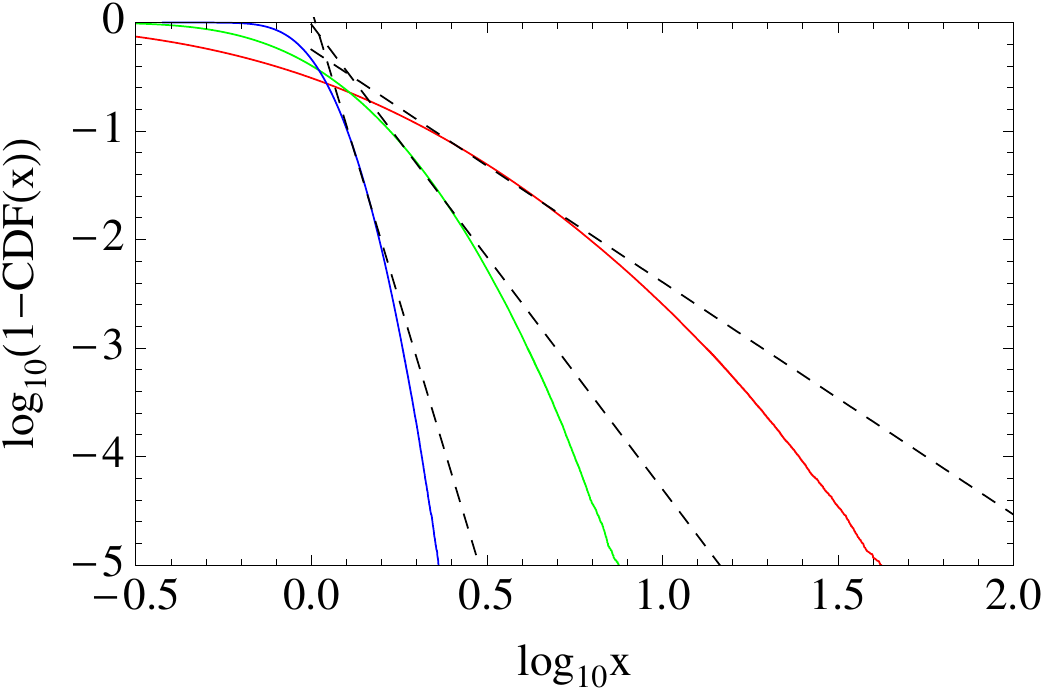}
\caption{Top: plots of PDF of $\LN(x; \mu,\si)$ with mean 1. The left red, middle green, and right blue curves correspond to parameter $\si = 1, 0.5$, and $0.2$ respectively. Bottom: log-log plots of simulated data sampled from the LN distributions. Below $-1$ of the y-axis, the left blue, middle green, and right red curves correspond to $\si = 0.2, 0.5$, and $1$ respectively. The dashed lines are fitting of $\log_{10}(1-\text{CDF}(x))$ vs. $\log_{10} x$ in a range of CDF from 0.9 to 0.99.}
\label{ST:fig:LN_simulation_loglogplot}
\includegraphics[width=0.345\textwidth]{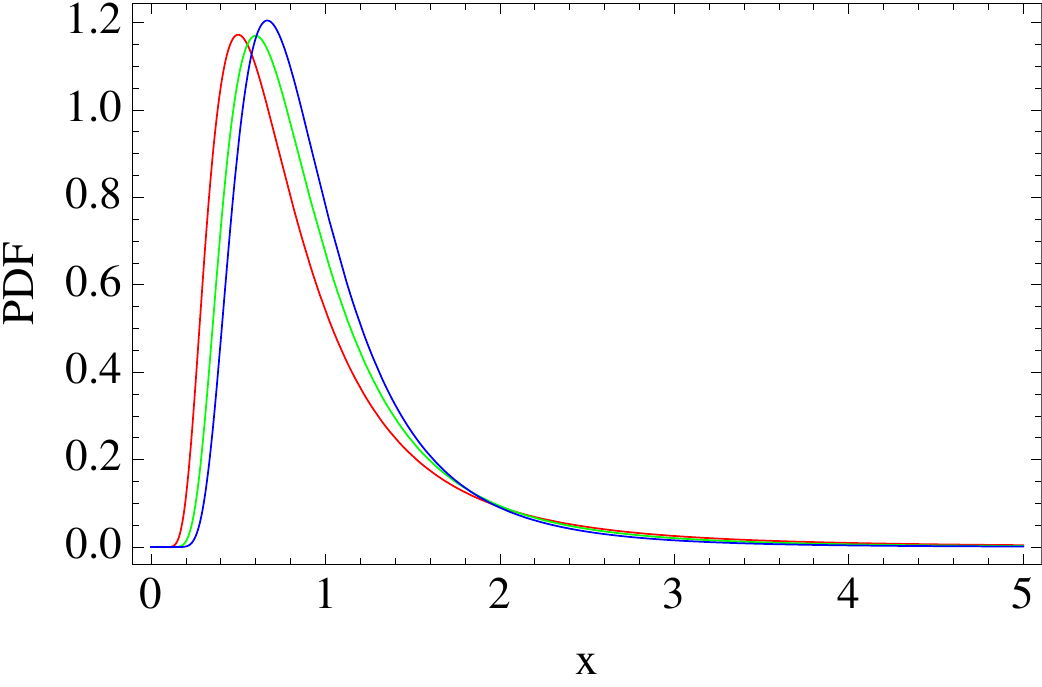}
\includegraphics[width=0.345\textwidth]{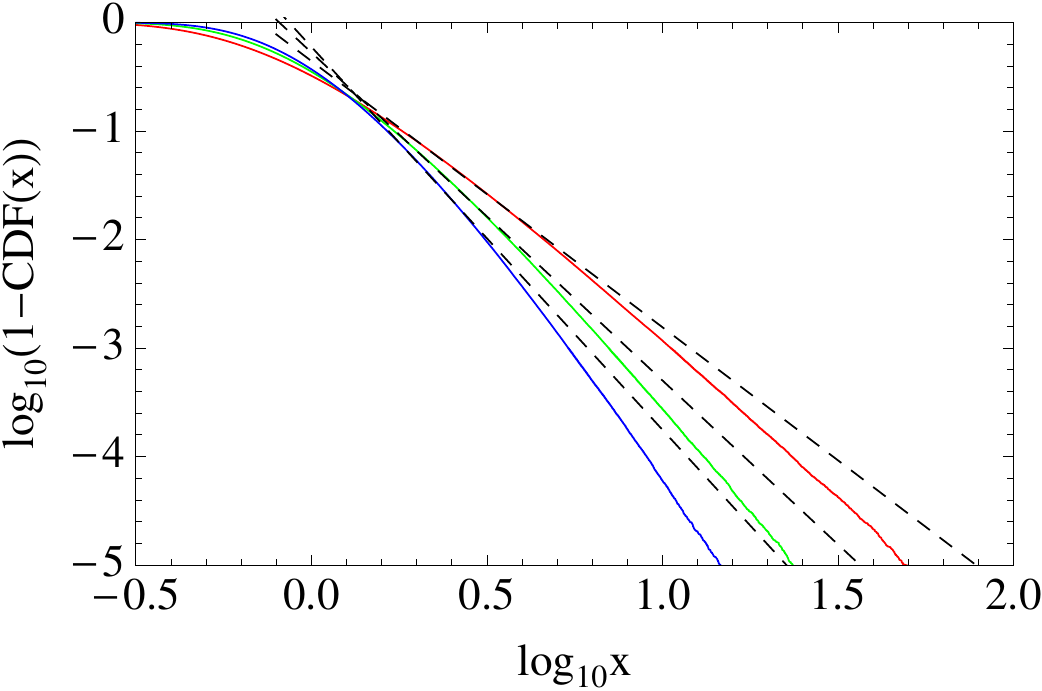}
\caption{Top: plots of PDF of $\IGa(x; \al,\be)$ with mean 1. The left red, middle green, and right blue curves correspond to $\al = 3, 4$, and $5$ respectively. Bottom: log-log plots of simulated data sampled from the IGa distributions. Below $-1$ of the y-axis, the left blue, middle green, and right red curves correspond to $\al = 5, 4$, and $3$ respectively. The dashed lines with slopes $-3.5, -3.0$, and $-2.5$ respectively are fitting of $\log_{10}(1-\text{CDF}(x))$ vs. $\log_{10} x$ in a range of CDF from 0.9 to 0.99. }
\label{ST:fig:IGa_simulation_loglogplot}
\end{figure}

\begin{figure}[htp]
\centering
\includegraphics[width=0.345\textwidth]{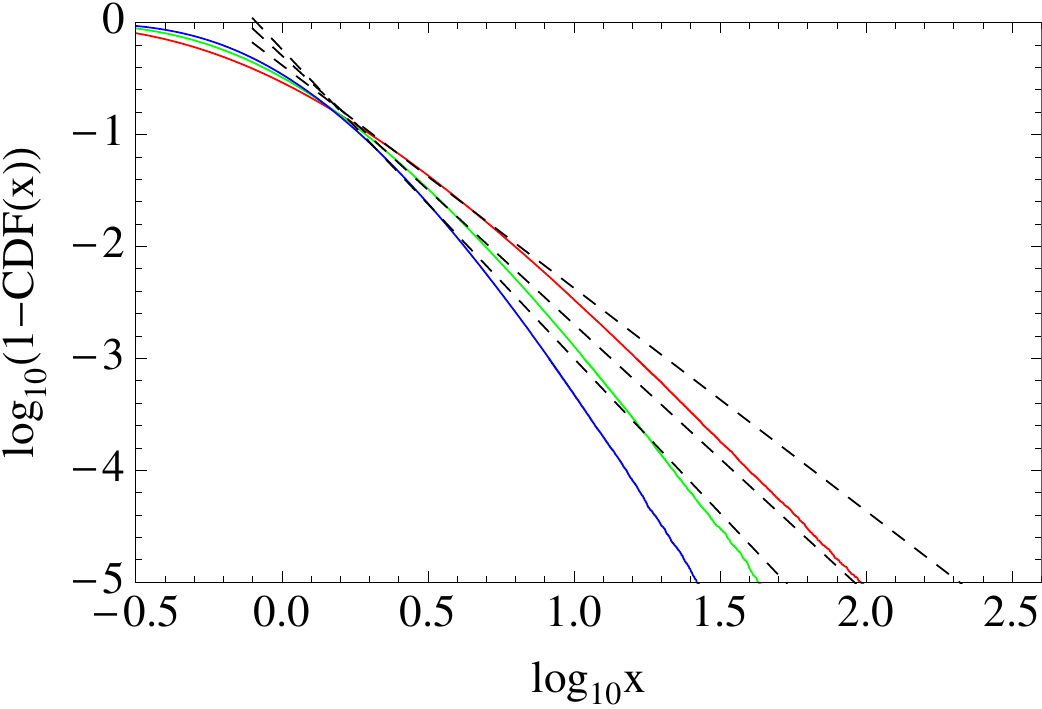}
\includegraphics[width=0.345\textwidth]{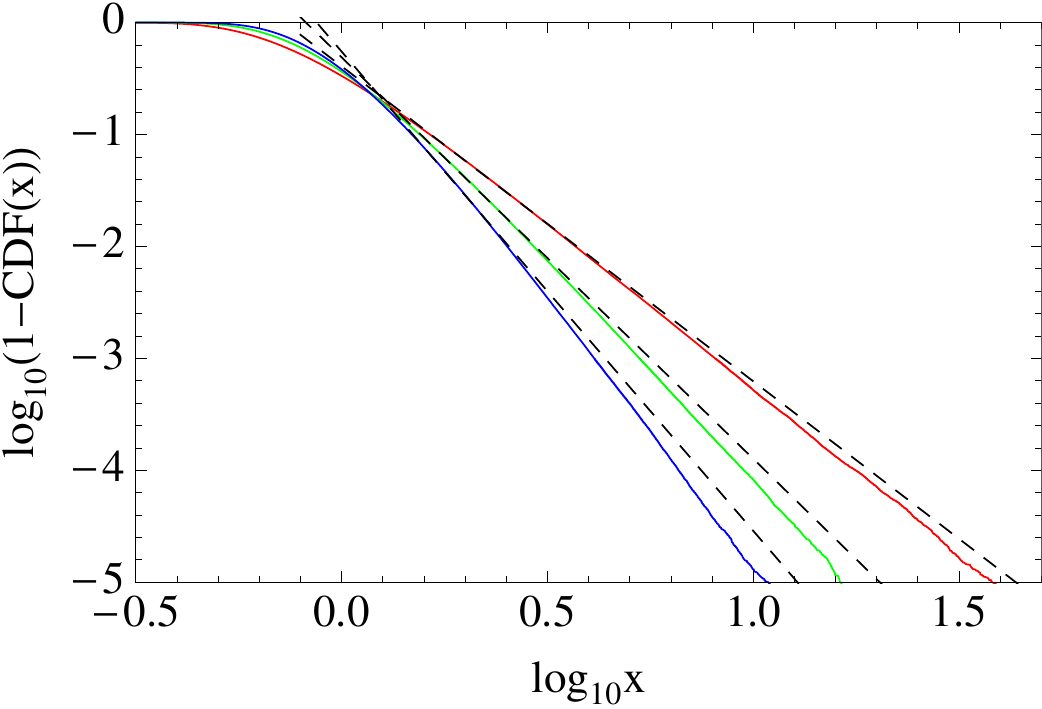}
\caption{Log-log plots of simulated data sampled from GIGa distributions $\GIGa(x; \al,\be, 0.5) (top)$ and $\GIGa(x; \al,\be,2)$ (bottom) with mean 1. Below $-1$ of the y-axis, the left blue, middle green, and right red curves correspond to $\al=2.5, 2$, and $1.5$ respectively. The dashed lines with slopes $-2.8, -2.4$, and $-2.0$ respectively (top) and $-4.3, -3.6$, and $-2.8$ (bottom) are fitting of $\log_{10}(1-\text{CDF}(x))$ vs. $\log_{10} x$ in a range of CDF from 0.9 to 0.99. }
\label{ST:fig:GIGa_ga_half_simulation_loglogplot}
\end{figure}

The exponent of a power law tail can be easily calculated once we notice that
\begin{equation}
1-\text{CDF}(x) = \int_x^{+\infty} \text{PDF}(x) dx . 
\end{equation}
If $\text{PDF}(x) \propto C x^{-1-\rho}$ with $x\gg 1$, then
\begin{equation}
\log( 1-\text{CDF}(x) ) \propto \text{const} - \rho \log x . 
\end{equation}
In Figs. \ref{ST:fig:LN_simulation_loglogplot} and \ref{ST:fig:IGa_simulation_loglogplot}, we show the log-log plot of the tail of LN and IGa distributions respectively. Clearly, a straight line fit is considerably better for the latter, even though the fitted slope does not coincide with the theoretical value. Towards this end, in Fig. \ref{ST:fig:GIGa_ga_half_simulation_loglogplot}, we show log-log plots of the tail of GIGa distributions for $\ga=0.5$ and $\ga=2$. The empirical trend emerging form the IGa and GIGa plots is that the straight line fits of log-log plots become progressively better as $\ga$ gets larger. 

\begin{figure}[htp]
\centering
\includegraphics[width=0.45\textwidth]{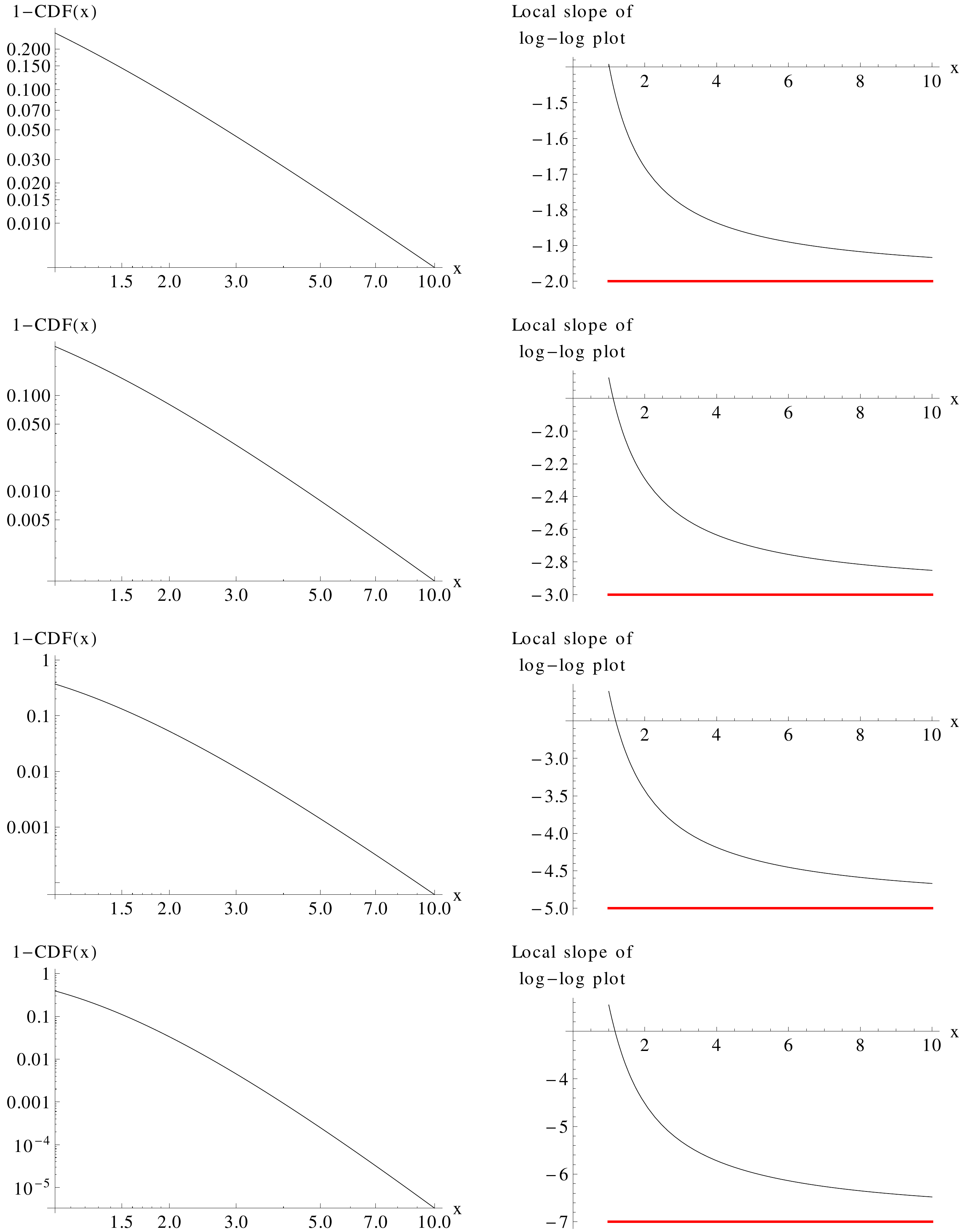}
\caption{Local slope of log-log plot of IGa distribution IGa$(x; \al, \be)$ with mean 1 ($\be=\Ga(\al)/\Ga(\al-1)$). The left column is the log-log plot and the right one is the local slope of the log-log plot from Eq. (\ref{RT:eq:local_slope_GIGa}). $\al$ is 2, 3, 5, and 7 for the first, second, third, and fourth rows respectively. The red lines are $-\al$: the limit of the local slope when $x\rightarrow\infty$.}
\label{RT:fig:local_slope_IGa}
\end{figure}

\begin{figure}[htp]
\centering
\includegraphics[width=0.45\textwidth]{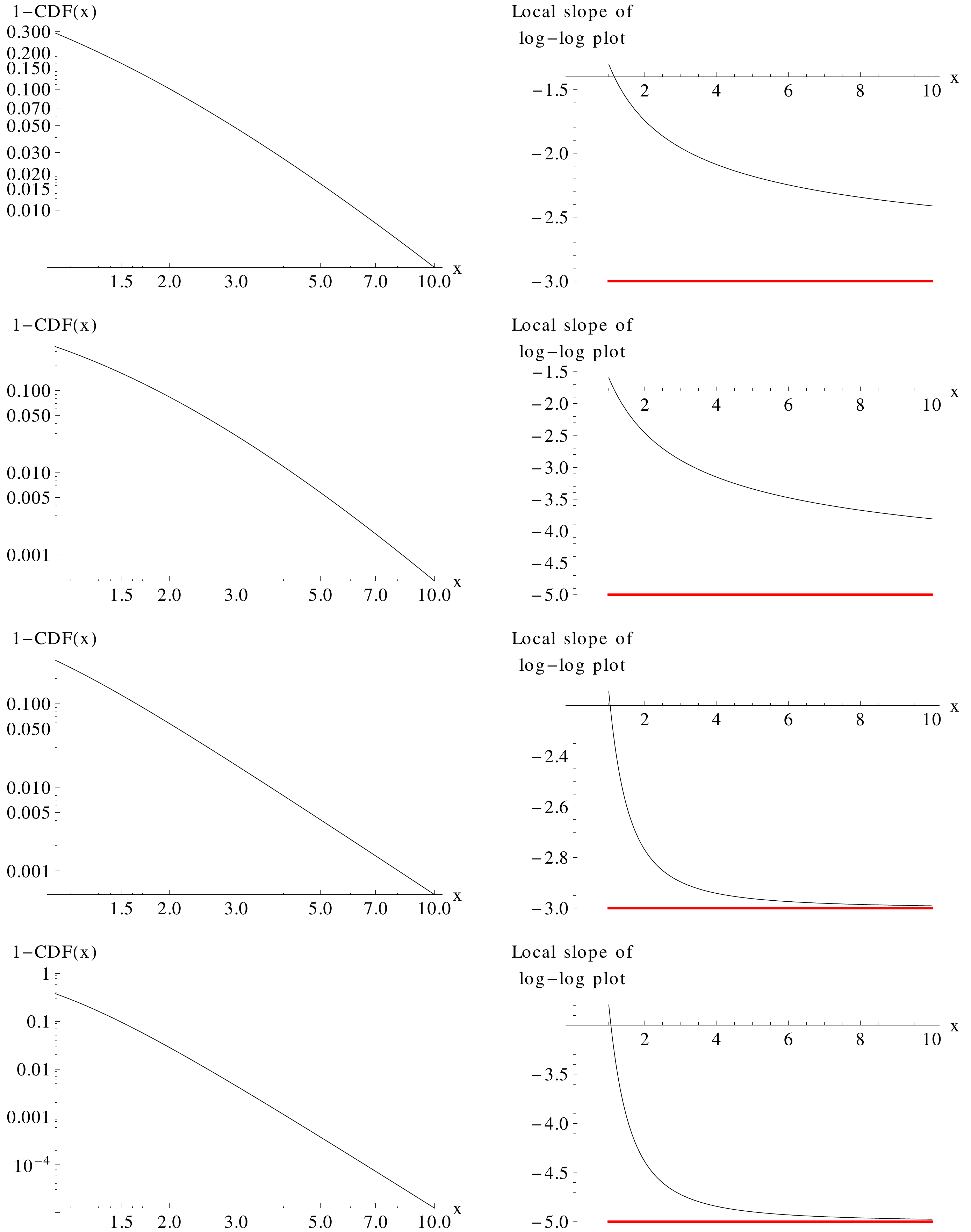}
\caption{Local slope of log-log plot of GIGa$(x; \al, \be, \ga)$ with mean 1 ($\be=\Ga(\al)/\Ga(\al-1/\ga)$). The left column is the log-log plot and the right one is the local slope of the log-log plot from Eq. (\ref{RT:eq:local_slope_GIGa}). $\{\al,\ga\}$ is $\{6,0.5\}$, $\{10,0.5\}$, $\{1.5,2\}$, and $\{2.5,2\}$ for the first, second, third, and fourth rows respectively. The red lines are $-\al\ga$: the limit of the local slope when $x\rightarrow\infty$.}
\label{RT:fig:local_slope_GIGa}
\end{figure}

To understand this $\ga$-dependence the difference between the theoretical and fitted slope, we consider the local slope of the log-log plot,
\begin{equation}\label{RT:eq:local_slope_def}
\fr{d\log(1-\text{CDF}(x))}{d\log x}.
\end{equation}
For GIGa (and IGa, $\ga=1$), the local slope is given by
\begin{equation}\label{RT:eq:local_slope_GIGa}
\fr{\ga  e^{-\lf(\fr{\be}{x}\rg)^{\ga}} \lf(\fr{\be}{x}\rg)^{\al\ga}}{\Gamma
   (\al) \lf(Q\lf(\al ,\lf(\fr{\be}{x}\rg)^{\ga}\rg)-1\rg)} ,
\end{equation}
with the regularized gamma function $Q(s,x)={\Ga(s,x)}/{\Ga(s)}$, where $\Ga(s,x)\equiv \int_x^\infty t^{s-1}e^{-t}dt$ is the incomplete gamma function. The local slopes are shown, as function of $x$ in Figs. \ref{RT:fig:local_slope_IGa} and \ref{RT:fig:local_slope_GIGa} respectively. It is clear that the local slope can differ substantially from its limiting (saturation) value. As $\ga$ becomes larger, the local slope tends closer to its limiting value.  

\begin{figure}[htp]
\centering
\includegraphics[width=0.45\textwidth]{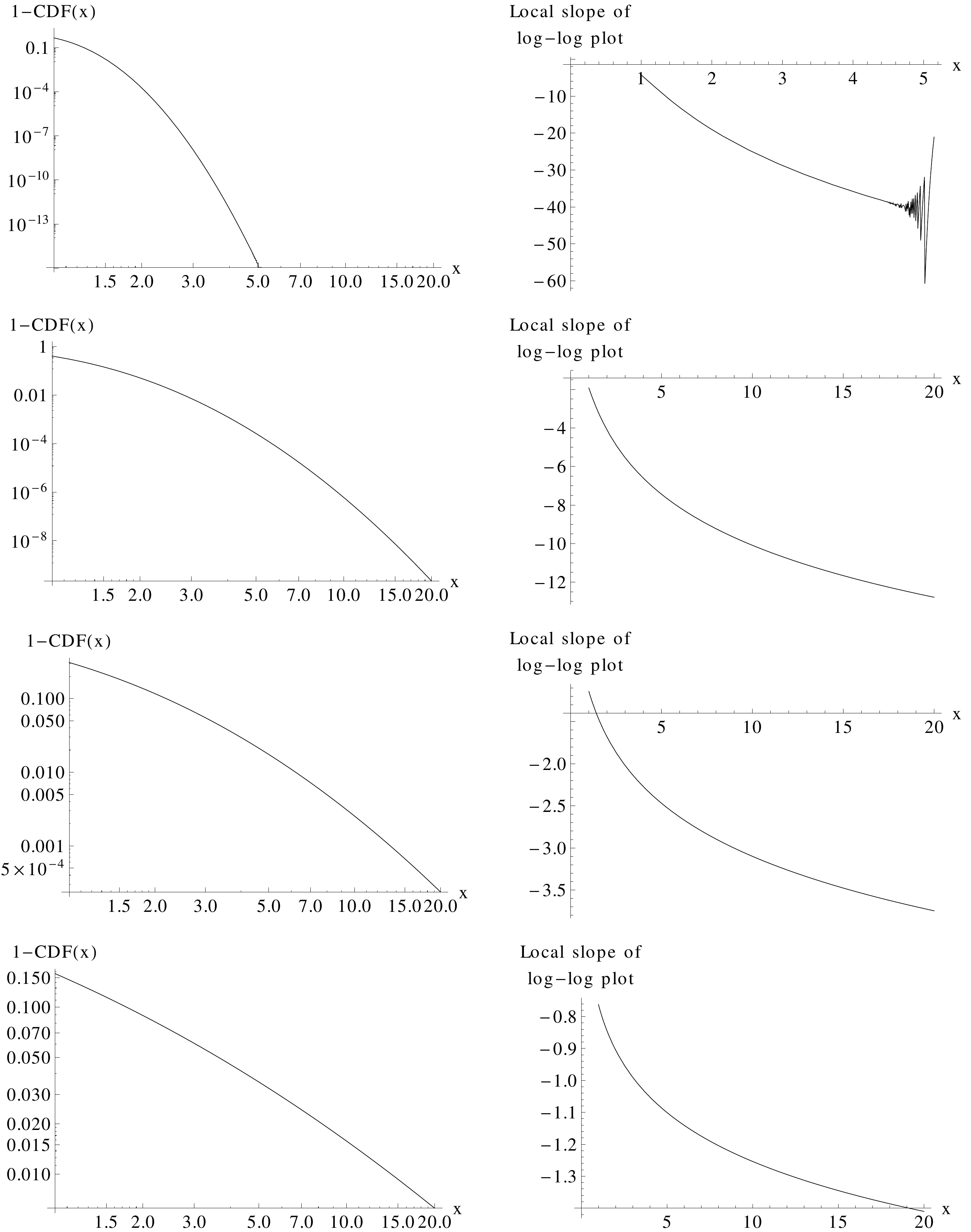}
\caption{Local slope of log-log plot of lognormal distribution. The mean of the distribution is set as 1 through $\mu = -\si^2/2$. The left column is the log-log plot and the right one is the local slope of the log-log plot in Eq. (\ref{RT:eq:local_scope_LN}). $\si$ is 0.2, 0.5, 1, and 2 for the first, second, third, and fourth row respectively. The jagged part of the top right plot is due to computational precision.}
\label{RT:fig:local_slope_LN}
\end{figure}

For the LN distribution, the local slope is given by
\begin{equation}\label{RT:eq:local_scope_LN}
\fr{\sqrt{\fr{2}{\pi}} e^{-\fr{(\log x  - \mu)^2}{2 \si ^2}}}{\si \lf(1 + \text{erf}\lf(-\fr{\log x - \mu}{\sqrt{2}\si}\rg)\rg)} ,
\end{equation}
which slowly decreases with $x$. But as is clear from (\ref{RT:eq:local_scope_LN}) and Fig. \ref{RT:fig:local_slope_LN}, the local slope does not saturate when $x\rightarrow \infty$. 

\bibliographystyle{unsrt}

\end{document}